\documentclass[a4paper,twocolumn,11pt,accepted=2021-06-28]{quantumarticle}
\pdfoutput=1
\usepackage[utf8]{inputenc}
\usepackage[english]{babel}
\usepackage[T1]{fontenc}
\usepackage{amsmath}
\usepackage{hyperref}

\usepackage{tikz}
\usepackage{lipsum}
\usepackage{amsmath}
\usepackage{wasysym}
\usepackage{mathtools}
\usepackage[numbers]{natbib}
\usepackage{nicefrac}
\usepackage{colortbl}
\makeatletter

\newcommand*\LyXThinSpace{\,\hspace{0pt}}
\pdfpageheight\paperheight
\pdfpagewidth\paperwidth

\newcommand{\epr}{\mathrm{EPR}}
\newcommand{\bel}{\mathrm{B-EPR}}
\newcommand{\sm}{Supplementary Information}
\usepackage{mathbbol}
\usepackage{tikz}

\newcommand{\up}{\textcircled{u}}
\newcommand{\dwn}{\textcircled{\raisebox{-.9pt}{d}}}

\newcommand{\cirp}{\textcircled{$+$}}
\newcommand{\cirm}{\textcircled{$-$}}
\usepackage[utf8]{inputenc}
\usepackage[T1]{fontenc}

\makeatother

\usepackage{babel}
\global\long\def\bk{\mathbf{k}}%

\global\long\def\epr{}%

\global\long\def\bel{\mathrm{B}}%
\begin{document}

\title{Real--time ghost imaging of Bell--nonlocal entanglement between a photon
and a quantum memory}
\author{Mateusz Mazelanik}
\affiliation{Centre for Quantum Optical Technologies, Centre of New Technologies,
University of Warsaw, Banacha 2c, 02-097 Warsaw, Poland}
\affiliation{Faculty of Physics, University of Warsaw, Pasteura 5, 02-093 Warsaw,
Poland}
\author{Adam Leszczyński}
\affiliation{Centre for Quantum Optical Technologies, Centre of New Technologies,
University of Warsaw, Banacha 2c, 02-097 Warsaw, Poland}
\affiliation{Faculty of Physics, University of Warsaw, Pasteura 5, 02-093 Warsaw,
Poland}
\author{Michał Lipka}
\affiliation{Centre for Quantum Optical Technologies, Centre of New Technologies,
University of Warsaw, Banacha 2c, 02-097 Warsaw, Poland}
\affiliation{Faculty of Physics, University of Warsaw, Pasteura 5, 02-093 Warsaw,
Poland}
\author{Wojciech Wasilewski}
\affiliation{Centre for Quantum Optical Technologies, Centre of New Technologies,
	University of Warsaw, Banacha 2c, 02-097 Warsaw, Poland}
\author{Michał Parniak}
\affiliation{Centre for Quantum Optical Technologies, Centre of New Technologies,
University of Warsaw, Banacha 2c, 02-097 Warsaw, Poland}
\affiliation{Niels Bohr Institute, University of Copenhagen, Blegdamsvej 17, DK-2100
Copenhagen, Denmark}
\email{m.parniak@cent.uw.edu.pl}

\maketitle

\begin{abstract}
Certification of nonlocality of quantum mechanics
is an important fundamental test that typically requires prolonged
data collection and is only revealed in an in-depth analysis. These
features are often particularly exposed in hybrid systems, such as
interfaces between light and atomic ensembles. Certification of entanglement
from images acquired with single-photon camera can mitigate this issue
by exploiting multiplexed photon generation. Here we demonstrate this
feature in a quantum memory (QM) operating in a real-time feedback mode.
Through spatially-multimode spin-wave storage the QM enables
operation of the real-time ghost imaging (GI) protocol. By properly preparing
the spatial phase of light emitted by the atoms we enable observation
of Bell-type nonlocality from a single image acquired in the far field
as witnessed by the Bell parameter of $S=2.227\pm0.007>2$. Our results
are an important step towards fast and efficient utilization of multimode
quantum memories both in protocols and in fundamental tests. 
\end{abstract}


Correlated photon pairs not only provide a basic tool for testing
the quantum theory, but also constitute a bedrock of modern quantum
optics where the applications as secure communication or quantum-enhanced
imaging are gathering a continuously rising interest. Starting from
seminal works of Clauser and Aspect \citep{Freedman1972,Aspect1982}
polarization-entangled photons have been used for decades to demonstrate
the quantum spookiness, namely by demonstration of the Bell-type correlations.
Tremendous efforts have been put to close numerous
loopholes \citep{Shalm2015,Giustina2015,Abellan2018,Vedovato2018}
 and to utilize different degrees of freedom (DoF) such as time-bin
\citep{Vedovato2018,Brendel1999}, position \citep{Yarnall2007} and
momentum \citep{Rarity1990} or angular momentum of photons \citep{Leach2009}
to trustfully rule out the local hidden variables theories by violation
of the Bell inequality \citep{Bell1964}. With the
development of experimental quantum optics, the Bell test emerged
as a versatile benchmarking tool providing a performance measure for
various quantum protocols \citep{Acin2006,Yuan2008}.

The Bell inequality \citep{Bell1964} arose as a response to work by Einstein, Podolsky and Rosen
(EPR) \citep{Einstein1935} concerning completeness of quantum theory.
The so-called EPR paradox suggesting the failure of local realism
has been initially demonstrated with quadratures of light \citep{Jensen2011,Ou1992}
and lately with position and momenta correlated photons \citep{Howell2004,Edgar2012,Moreau2014,Dabrowski2017}. The quantum ghost imaging (GI) \citep{Strekalov1995,Pittman1995} - a
technique that uses correlated photon pairs to reconstruct an image
from photons that do not interact with the object being imaged - emerged
from the latter.

In the modern quantum GI protocol the object being imaged
is placed in either near or far field of an EPR photon source. The
source generates correlated photons in two distinct beams - a \emph{signal}
and \emph{idler}. The object is illuminated by the \emph{signal} photons
which are then detected using non-spatially-resolving \emph{bucket
}detector acting as a trigger source for second, spatially-resolving
detector - a camera placed in the \emph{idler} beam in the same optical
plane as the object. As the camera registers \emph{idler} photons
only when a \emph{signal} photon is transmitted through the object
(a trigger signal is generated) the \emph{idler} photon events build
up the image. In such scenario the \emph{idler} photons have to be
additionally time-delayed to surpass the trigger signal electronic
propagation delay, which was up to date only achieved via an
image-preserving optical delay line \citep{Moreau2019a}, while in
other cases researchers relied on post-selection of correlated events.
On the other hand, a spatially-multimode QM \citep{Parniak2017}
is a natural choice when one needs to store the idler photon and
release it upon the trigger signal, thus gaining more versatility,
potentially much larger storage times and simultaneously avoiding
difficult engineering of delay lines. 

Here we employ a wavevector-multiplexed emissive QM
\citep{Parniak2017} to demonstrate a real-time GI of polarization
and wavevector DoF Bell-type correlations. We utilize our ability
of EPR-type entanglement generation \citep{Dabrowski2018} to prepare
polarization Bell-states in many wave-vector modes. The tens of $\mu s$-long
storage time of our memory enables us to easily realize a heralded
GI setup without any additional optical delay line. The
images of Bell-type correlations are acquired using a state-of-the
art single photon sensitive camera giving almost 79\% visibility and
violation of the Bell inequality by 32 standard deviations. With that, we bring this extremally versatile approach to Bell inequalities to a hybrid atom-photon system. This is not only of fundamental interest, but the atomic medium brings about a set of possible manipulations \cite{Parniak2019} and interactions \cite{Peyronel2012}.

The role of quantumness in GI has been a topic of debate
for past two decades. While it is currently known that the ghost images
can be produced with classical light \citep{Bennink2002,Valencia2005}
the quantum GI can lead to improved contrast or signal
to noise ratio (SNR) \citep{Aspden2016,Jack2009} and resolution \citep{Dangelo2005}
in the produced image. Therefore, the current question is not the
quantum nature of GI but rather if with the quantum illumination
a new features can be observed. Finally, the recent demonstrations
of imaging of Bell-type nonlocal behavior \citep{Moreau2019} and
quantum GI utilizing higher order correlations \citep{Hodgman2019}
suggest a sparking interest in this topic, with our work bringing
this issue of fundamental interest into a hybrid atom-photon system.


\begin{figure}[t]
\includegraphics[width=1\columnwidth]{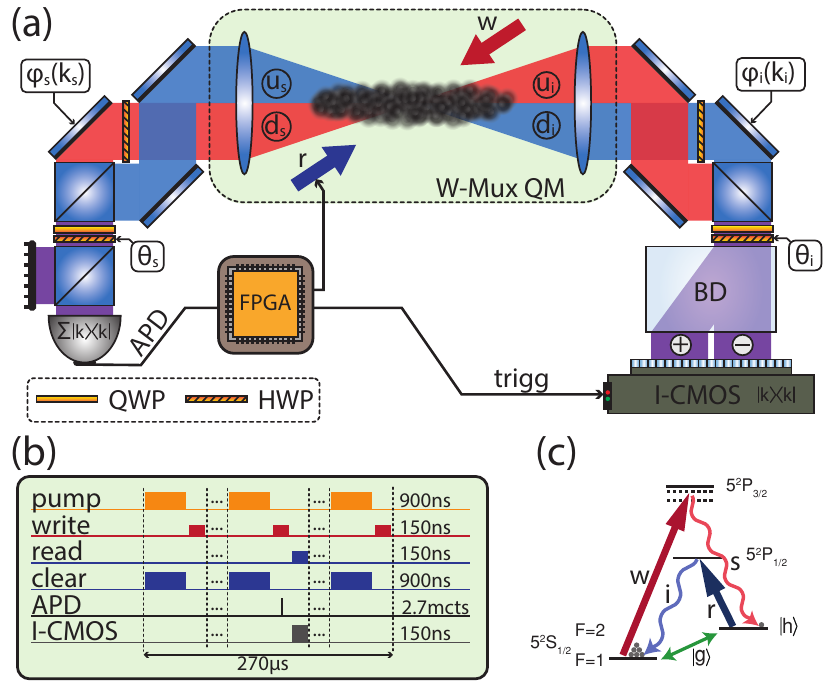}

\caption{Experimental setup for QM-assisted phase sensitive
GI. (a) The wavevector multiplexed quantum memory W-Mux
QM generates time delayed EPR\emph{ signal} and \emph{idler} pairs.
The Mach-Zehnder interferometers (MZI) superimposes two halfs of the
wavevetor modes onto each other generating wavevector indexed superposition
of polarization DoF Bell-states. The MZIs provides also additional
control of the Bell-state phase $\varphi_{s}-\varphi_{i}$. The \emph{signal}
photons are detected by \emph{bucket }detector placed after polaryzer
at angle $\theta_{s}$. The \emph{signal} detection events are registered
by FPGA controlling the experimental sequence (especially the readout
pulse r) and providing a trigger for the I-CMOS camera. (b) The experimental
sequence. The readout is performed only if a \emph{signal} photon
is detected. (c) Relevant $\mathrm{^{87}Rb}$ energy levels used in
the QM protocol.\label{fig:Experimental-setup}}
\end{figure}

We use wavevector multiplexed multimode QM
as a quantum light source for phase sensitive GI \citep{Parniak2017}.
Our experimental setup is shown in Fig. \ref{fig:Experimental-setup}(a) and described in more detail in the \sm.
In the memory, the signal photons are generated together with spin
waves. After a programmable delay, the spin wave is converted into
idler photons. The resulting photon pair is correlated in positions
and anti-correlated in momenta (EPR state), while the polarizations
are uncorrelated and simply set by collection optics. Therefore, we
may write the resulting two-photon state as: 
\begin{equation}
|\psi_{\epr}\rangle=\int\psi_{\epr}(\mathbf{k}_{s},\mathbf{k}_{i})|\mathbf{k}_{s},H\rangle|\mathbf{k}_{i},H\rangle d\mathbf{k}_{s}\mathrm{d}\mathbf{k}_{i},\label{eq:psik}
\end{equation}
where $H$ stands for horizontal polarization and the biphoton wavefunction
$\psi_{\mathrm{\epr}}$ in the realistic scenario can be approximated
by \citep{Dabrowski2018,Parniak2017,Edgar2012}: 
\begin{equation}
	\psi_{\epr}(\mathbf{k}_{s},\mathbf{k}_{i})=\mathcal{\frac{\sigma}{\pi\kappa}}e^{-\frac{(\mathbf{k}_{s}+\mathbf{k}_{i})^{2}}{4\kappa^{2}}-\frac{\sigma^{2}(\mathbf{k}_{s}-\mathbf{k}_{i})^{2}}{4}},\label{eq:wf}
\end{equation}
where the Gaussian widths $\kappa$ and $\sigma^{-1}$ corresponds
respectively to strength of the momenta and position correlations.
Furthermore, as we are limited by the numerical aperture rather than
spread of the phase matching spectrum determining the set of available
emission angles, we will assume perfect correlation in position and
take $\sigma\to0$. 

The \emph{signal} and \emph{idler} photons are collected by two identical
far-field imaging setups (represented as a single lenses on Fig. \ref{fig:Experimental-setup}(a)). Next, the photons
pass through setups similar to MZI,
where each of the beams (either \emph{signal} or \emph{idler}) is
split in half and resulting components (upper-\up\  and lower-\dwn
) are joined together on the polarizing beam splitter (PBS), with
a help of a half-wave plate. Additionally, as the MZIs are placed
in the far-filed of the ensemble, tilting one of the MZI mirrors results
in a wavevector-dependent phase between the two joined paths (see
Fig. \ref{fig:Experimental-setup}(a)). The resulting state can be
written as: 
\begin{widetext}
\begin{gather}
|\psi_{\bel}\rangle=
\mathcal{N}\iint_{[0,\delta\bk]}e^{i\varphi_{s}(\mathbf{k}_{s})}\psi_{\epr}(\mathbf{k}_{s},\mathbf{k}_{i}-\delta\mathbf{k})|\mathbf{k}_{s},H\rangle|\mathbf{k}_{i},V\rangle+e^{i\varphi_{i}(\mathbf{k}_{i})}\psi_{\epr}(\mathbf{k}_{s}-\delta\mathbf{k},\mathbf{k}_{i})|\mathbf{k}_{s},V\rangle|\mathbf{k}_{i},H\rangle)\mathrm{d}\mathbf{k}_{s}\mathrm{d}\mathbf{k}_{i}\equiv\nonumber \\
\iint_{[0,\delta\bk]}\tilde{\psi}_{\epr}(\mathbf{k}_{s},\mathbf{k}_{i})|\mathbf{k}_{s}\rangle|\mathbf{k}_{i}\rangle(|H\rangle|V\rangle+e^{i\varphi_{s}(\mathbf{k}_{s})-i\varphi_{i}(\mathbf{k}_{i})}|V\rangle|H\rangle)\mathrm{d}\mathbf{k}_{s}\mathrm{d}\mathbf{k}_{i},\label{eq:psi_bell}
\end{gather}
\end{widetext}
where $\mathcal{N}$ is the normalization constant, $\tilde{\psi}_{\epr}(\mathbf{k}_{s},\mathbf{k}_{i})$
is the renormalized $\psi_{\epr}(\mathbf{k}_{s}-\delta\bk/2,\mathbf{k}_{i}-\delta\bk/2)$
and $\delta\mathbf{k}=\delta k\hat{\mathbf{y}}$ with $\delta k=286\,\mathrm{mm^{-1}}$
is the wavevector shift applied to the upper (\up\  on Fig. \ref{fig:Experimental-setup}(a))
part of beam to superimpose it on the lower part (\dwn\  on Fig.
\ref{fig:Experimental-setup}(a)). By $\varphi_{s(i)}(\mathbf{k}_{s(i)})$
we denote the additional phase added before the superimposition at
each side. The $|\psi_{\mathrm{\bel}}\rangle$ represents a wavevector-indexed
superposition of polarization DoF Bell-states and exhibits both EPR
and Bell-type correlations. 

\begin{figure}[t]
\includegraphics[width=1\columnwidth]{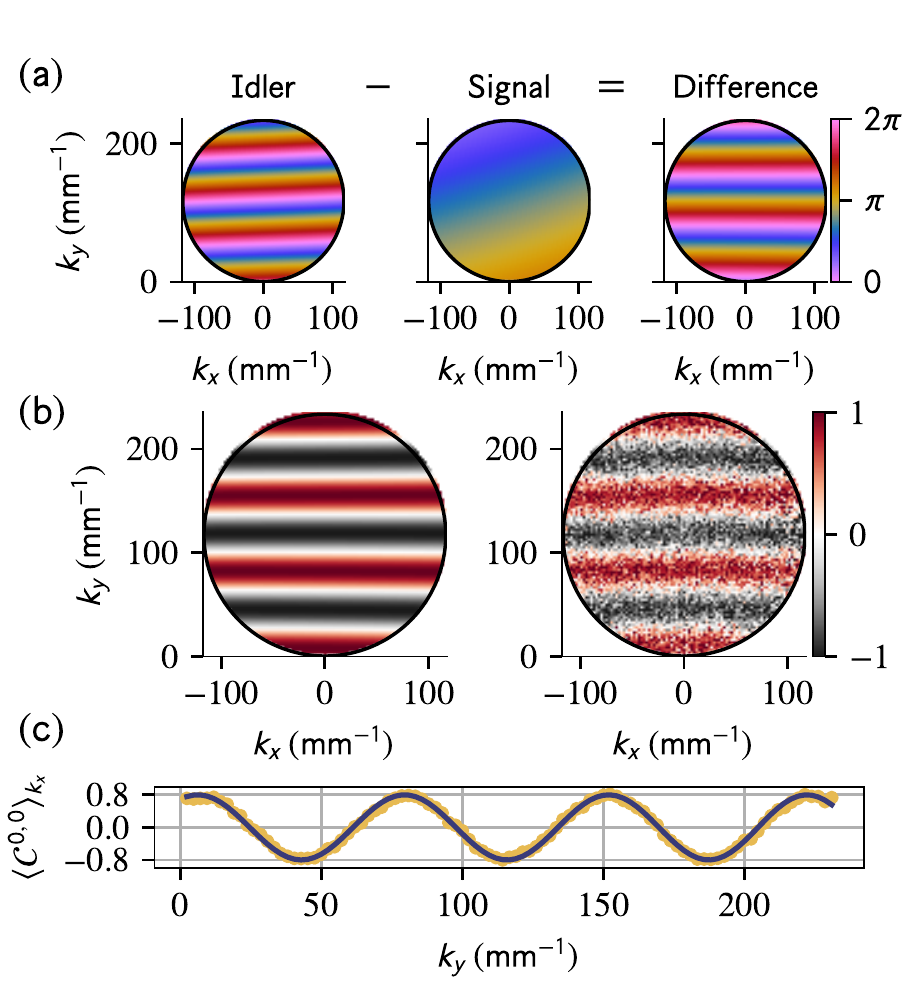}

\caption{Phase profiles for Bell entanglement preparation and verification.
(a) Measured phase profiles in the \emph{idler} and \emph{signal}
arms and the resulting phase difference profile relevant for the generated
state. (b) Comparison between the classically simulated (left) and
measured (right) correlation function $\mathcal{C}^{0,0}(\protect\bk_{i})$.
(c) The $k_{x}$--averaged fringes from (b): line--simulation, points--measurement.
\label{fig:phase}}
\end{figure}

To exploit the latter we perform a polarization measurement over large
set of \emph{signal} and \emph{idler} wavevectors using a GI
technique with feedback enabled by the QM feature of our
source. In contrast to ``single-channel'' \citep{Aspect1981} or
``dual-channel'' \citep{Aspect1982} experiments we implement a
hybrid scenario in which one party (\emph{signal}) measures using
a ``single channel'' polarizer and a \emph{bucket} detector when
the second one (\emph{idler}) uses a ``dual-channel''
polarizer (BD on Fig. \ref{fig:Experimental-setup}(a)) and single-photon-sensitive
camera (I-CMOS) \cite{doi:10.1063/1.5033559} placed in the far-field of the atomic
ensemble. The \emph{bucket} detector [a multimode-fiber-coupled
avalanche photiodiode (APD)] is connected to the experimental sequence
controller (FPGA), which decides whether to perform the readout and gate
the camera or not (experimental sequence in Fig. \ref{fig:Experimental-setup}(b)).
This way the readout process, and most importantly, the camera measurement
is performed only if a \emph{signal} photon is detected. This feedback is the central point of our experimental implementation and provides a performance boost compared with hitherto post-selection approaches.

The joint measurement of \emph{signal}-\emph{idler} pairs can be described
by wavevector-indexed set of POVMs:
\begin{equation}
\Pi_{\pm}^{\theta_{s},\theta_{i}}(\bk_{i})=\int|\bk,\theta_{s}\rangle\langle\bk,\theta_{s}|\otimes|\bk_{i},\pm\theta_{i}\rangle\langle\bk_{i},\pm\theta_{i}|\mathrm{d}\mathbf{k},\label{eq:povm}
\end{equation}

where $|\theta\rangle=(|H\rangle+e^{i\theta}|V\rangle)/2$ represents
the measurement setting on the Bloch sphere's equator and by $\pm$
we denote the two possible outcomes on the ``dual-channel'' polarizer.
Then, the outcome probability distribution is calculated as:
\begin{gather}
p_{\pm}^{\theta_{s},\theta_{i}}(\bk_{i})=\langle\psi_{\bel}|\Pi_{\pm}^{\theta_{s},\theta_{i}}(\bk_{i})|\psi_{\bel}\rangle\nonumber \\
=\int|\tilde{\psi}_{\epr}(\bk_{s},\bk_{i})|^{2}\cos^{2}\left(\frac{\phi(\bk_{s},\bk_{i})+\theta_{s}\mp\theta_{i}}{2}\right)\mathrm{d}\bk_{s}\label{eq:prob}
\end{gather}
where $\phi(\bk_{s},\bk_{i})\coloneqq\varphi_{s}(\bk_{s})-\varphi_{i}(\bk_{i})$
is the phase difference profile and the integration goes over the
\emph{bucket} detector area, which in our case is a circle bounded
by the multimode fiber numerical aperture (NA=0.2). From Eq. \eqref{eq:prob}
we see that the resolution of the phase-phase sensitive GI
is limited by the strength of the wavevector correlation described
by Eq. \eqref{eq:wf}, i.e the phase change in the \emph{signal} arm
$\varphi_{s}(\bk_{s})$ should be slow when compared to anti-correlation
width $\kappa\approx6\,\mathrm{mm^{-1}}$ (see \sm). 
\begin{figure*}[t]
\includegraphics[width=1\textwidth]{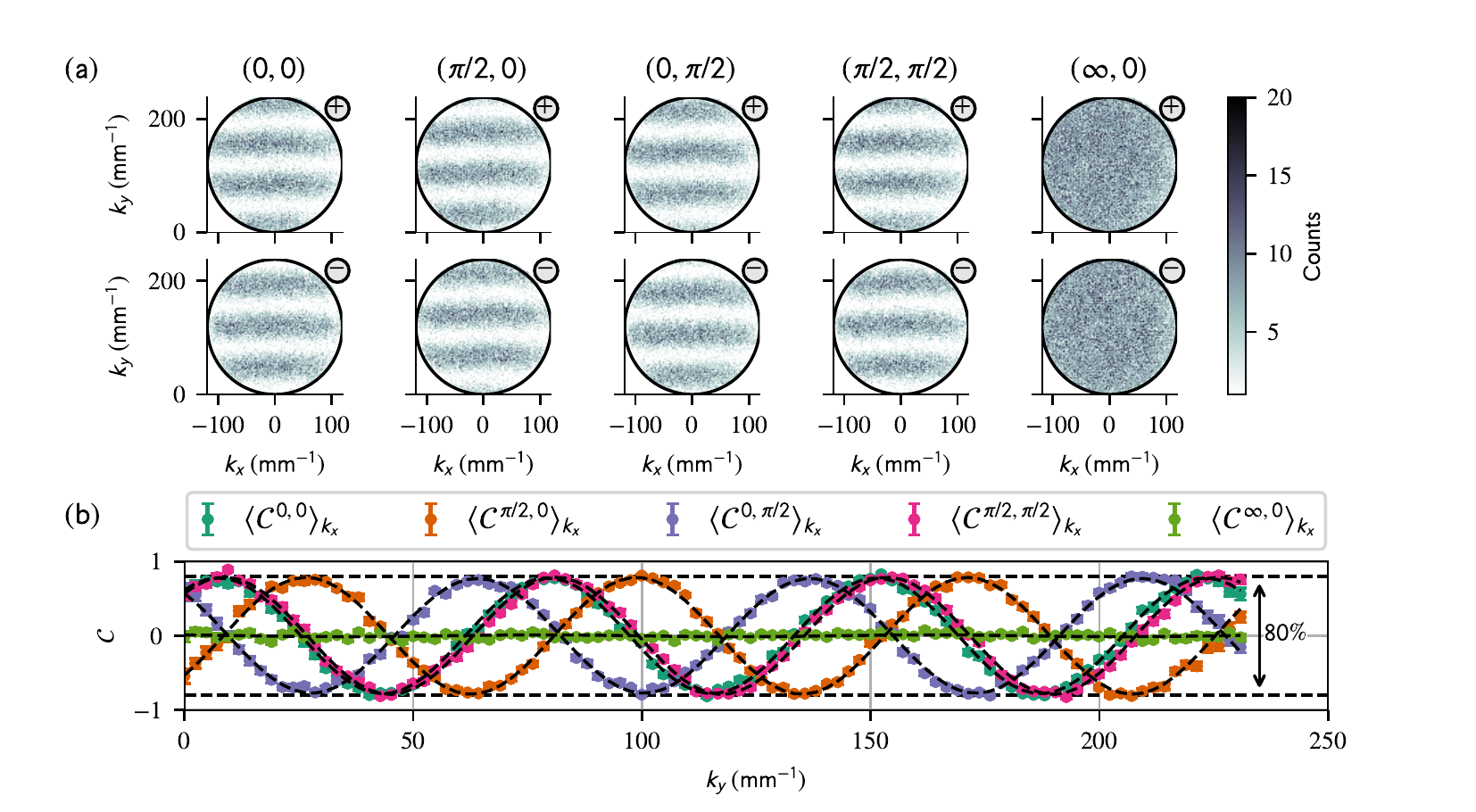}

\caption{Ghost images for different measurement settings in Bell state
characterization. (a) Ghost images of \emph{signal}-\emph{idler}
correlations for five $(\theta_{s},\theta_{i})$ measurement settings.
The \cirp\  and \cirm\  mark regions corresponding to two channels
($\pm$) of the ``dual channel'' polarizer. (b) Horizontal ($k_{x}$)
averages of the correlation function $\mathcal{C}^{\theta_{s},\theta_{i}}(k_{x},k_{y})$
for each measurement setting from (a). The local visiblity is up to
$80\%$. \label{fig:corrs}}
\end{figure*}

In our setup we choose the $\varphi_{s}(\bk_{s})$ to be linearly
and slowly varying mostly along the $y$ direction and similarly for
$\varphi_{i}(\bk_{i})$ but with a higher slope. To measure the phase
profiles in the \emph{signal} and \emph{idler} arms we classically
simulate the outcome by seeding the memory with coherent state, using
an additional laser beam focused at the atomic cloud centre (see \sm\
and \citep{Parniak2019}). For this measurement we replace the \emph{bucket}
detector with a camera and calcite beam displacer (see
\sm\  for details), and register intensity fringes in both arms. We
collect frames for various measurement settings $\{\theta_{s},\theta_{i}\}\in[0,2\pi]\times[0,2\pi]$
then for each frame we retrieve the phase using standard Fourier-transform
based procedure \citep{Lipka2019}, and finally average the results.
The retreived phase profiles are presented in Fig. \ref{fig:phase}(a),
where we also plot the combined profile corresponding to the $\phi(\bk_{s},\bk_{i})$
phase difference profile. Figure~\ref{fig:phase}(b) presents the
comparison between the classically simulated (from retrieved phase)
and measured correlation function $\mathcal{C}^{0,0}(\bk_{i})$ defined
as:

\begin{equation}
\mathcal{C}^{\theta_{s},\theta_{i}}(\bk_{i})=\frac{n_{+}^{\theta_{s},\theta_{i}}(\bk_{i})-n_{-}^{\theta_{s},\theta_{i}}(\bk_{i})}{n_{+}^{\theta_{s},\theta_{i}}(\bk_{i})+n_{-}^{\theta_{s},\theta_{i}}(\bk_{i})},\label{eq:corrf}
\end{equation}
where by $n_{\pm}^{\theta_{s},\theta_{i}}(\bk_{i})$ we denote the
number of registered idler photons with $\bk_{i}$ wavevector in each
port of the ``dual channel'' polarizer. From Eq. \eqref{eq:prob}
we expect the correlation function to be proportional to $\cos(\phi(-\bk_{s},\bk_{i}))$,
and this is indeed what we see. The $\mathcal{C}^{0,0}(\bk_{i})$
map is build from $2\times10^7$ experiment repetitions and contain $3\times10^{5}$
counts corresponding to \emph{signal}-\emph{idler} coincidences. In
Fig.~\ref{fig:phase}(c) we plot $k_{x}$-averaged fringes for
both maps where we see good agreement between the expectation and
measurement. The only parameter fitted there is the proportionality
constant interpreted as the visibility $\mathcal{V}=78\%$.


To perform the Bell test in our hybrid scenario we acquire
ghost images for four different measurement settings $(\theta_{s},\theta_{i})\in\{(0,0),(\nicefrac{\pi}{2},0),(0,\nicefrac{\pi}{2}),(\nicefrac{\pi}{2},\nicefrac{\pi}{2})\}\eqqcolon\mathcal{M}$
and for one additional (marginal) setting denoted as $\{\infty,0\}$
with the the single channel polarizer removed. We use a variation
of the CHSH inequality \citep{Clauser1969,Clauser1974,Aspect1982}:
\begin{equation}
|S|\leq2,\label{eq:CHSH}
\end{equation}
with the Bell parameter $S$ defined as follows (see \sm\ for derivation): 
\begin{multline}
S=\mathcal{C}^{\theta_{s},\theta_{i}}(\bk_{i})-\mathcal{C}^{\theta_{s},\theta_{i}^{\prime}}(\bk_{i})+\mathcal{C}^{\theta_{s}^{\prime},\theta_{i}}(\bk_{i})+\mathcal{C}^{\theta_{s}^{\prime},\theta_{i}^{\prime}}(\bk_{i})\\
-2\mathcal{C}^{\infty,\theta_{i}}(\bk_{i}).\label{eq:eS}
\end{multline}
Thanks to the wavevector dependent phase difference $\phi(\bk_{s},\bk_{i})$
that in our case changes more than $6\pi$ over observation region,
the $|\psi_{\bel}\rangle$ state (Eq. \eqref{eq:psi_bell}) maximally
violates the inequality \eqref{eq:CHSH} for any two pairs of measurement
settings separated by $\pi/2$ : $|\theta_{s}-\theta_{s}^{\prime}|=|\theta_{i}-\theta_{i}^{\prime}|=\pi/2$.
It stems from the fact that there is always a particular $\bk_{i}$
and thus $\phi(\bk_{s},\bk_{i})$ for which these settings are optimal,
i.e. lead to maximal violation of inequality \eqref{eq:CHSH}.
Therefore, for the sake of simplicity we choose the angles to be $0$
or $\pi/2$.

In Fig. \ref{fig:corrs}(a) we present acquired ghost images for the
five particular measurement settings, for both polarizer channels
denoted as $+$ and $-$. From these images we calculate the correlation
functions as defined by Eq. \eqref{eq:corrf}. The $k_{x}$-averaged
results are presented in Fig. \ref{fig:corrs}(b). From these, by
fitting the cosine function we obtain the following visibilities \{$77.9\%,78.6\%,77.2\%,78.9\%\}\pm0.5\%$
 corresponding to the four measurement settings from $\mathcal{M}.$
For the marginal setting (with the ``single channel'' polarizer
removed) we obtain $\mathcal{C}^{\infty,\theta_{i}}(\bk_{i})=(1\pm3)\times10^{-3}$
that together with the visibilities give $S=2.213\pm0.008$. This
violates the CHSH inequality \eqref{eq:CHSH} by 26 standard deviations
(SDs), indicating the quantum nature of the observed fringes. 
The less than 100\% visibility is a result of some imperfections present
at different stages of our setup, which are discussed in detail in the \sm.

To directly demonstrate the high visibility of the correlation function
for the four measurement bases we introduced the wavevector-dependent
phase difference profile $\phi(\bk_{s},\bk_{i})$ in the generated
state. Alternatively, as this phase is built from two independent
phase profiles in \emph{signal} and \emph{idler} arms we, could interpret
it as to be responsible for different measurement settings for different
wavevectors $\bk_{i}$. This way, by treating the MZIs with the known
phase profiles $\varphi_{s(i)}(\bk_{i})$ as a parts of the measurement
devices we can run the Bell test on the wavevector DoF entangled state
given by Eq.\eqref{eq:psik}. In this way, we register many independent $(\varphi_{s},\varphi_{i})$ 
measurement settings on a single
ghost image. Consequently, it is convenient to redefine the Bell $S$
parameter using the Freedman and Clauser formula \citep{Freedman1972}
modified for our hybrid scenario (see \sm\ for derivation): 
\begin{equation}
S=3\mathcal{C}(\phi)-\mathcal{C}(3\phi)-2\mathcal{C}^{\infty}(\phi),\label{eq:single_S}
\end{equation}
where $\mathcal{C}(\phi=\varphi_{s}-\varphi_{i})$ is defined using Eq.~\eqref{eq:corrf} with $n_{\pm}^{\theta_{s},\theta_{i}}(\bk_{i})$
averaged over the wavevectors $\bk_{i}$ corresponding to the same
phase $\phi$: $n_{\pm}^{\theta_{s},\theta_{i}}(\bk_{i})\to n_{\pm}^{\theta_{s},\theta_{i}}(\phi)=\langle n_{\pm}^{\theta_{s},\theta_{i}}(\bk_{i})\rangle_{\bk_{i}:\phi(-\bk_{i},\bk_{i})=\phi}$.
The marginal correlation function $\mathcal{C}^{\infty}(\alpha)$
is defined in the same manner and is experimentally found to be independent
of $\bk_{i}$ and thus $\phi$ as can be seen in Fig. \ref{fig:corrs}(b). 

\begin{figure}[t]
\includegraphics[width=1\columnwidth]{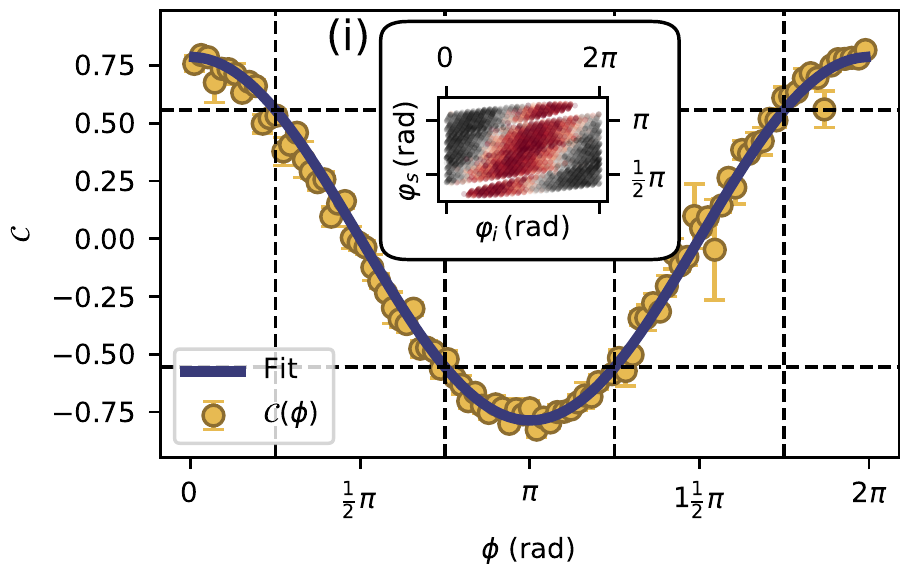}

\caption{Entanglement certification with a singule ghost image. Correlation
function $\mathcal{C}(\phi)$ from single ghost image ($(0,0)$ settings
from Fig.\ref{fig:corrs}(a)) along with fitted theoretical curve.
Inset (i) shows the ensemble of angular samples colored according
to the correlation function value for each point, similarly as in
Fig. \ref{fig:phase}(b). \label{fig:single_corr}}
\end{figure}

Figure~\ref{fig:single_corr} presents values of $\mathcal{C}(\phi)$
correlation function obtained from single ghost image (the $(0,0)$
setting of Fig.~\ref{fig:corrs}(a)). The ensemble of \emph{signal}
and \emph{idler} phase samples used to calculate $\mathcal{C}(\phi)$
is drawn in the inset (Fig.~\ref{fig:single_corr}(i)) where the coloring,
similarly to Fig.~\ref{fig:phase}(b), represents the value of the
correlation function for each $(\varphi_{s},\varphi_{i})$.
We clearly see that the correlation varies with the phase difference
$\phi=\varphi_{s}-\varphi_{i}$ as expected. With the definition of
the Bell parameter given by Eq.~\eqref{eq:single_S} the inequality
Eq.\eqref{eq:CHSH} is maximally violated for four phases from the
$[0,2\pi]$ range: $\phi\in\{\pi/4$, $3\pi/4,5\pi/4,7\pi/4$\}. Those
points and the corresponding values of $\mathcal{C}(\phi$) are marked
in the Fig. \ref{fig:single_corr} using dashed lines. 

By fitting the visibility $\mathcal{V}=78.6\%\pm0.3\%$
of the expected correlation curve $\mathcal{V}\cos(\phi)$ to match
the observed correlation values $\mathcal{C}(\phi)$ we obtain $S=2.227\pm0.007$
which violates the inequality Eq.\eqref{eq:CHSH} by more than 32
SDs.


In summary, we have employed an emissive QM as a quantum light source for
phase-sensitive GI setup and demonstrated images of Bell-type
correlations. The QM generates entangled \emph{signal }and\emph{
idler} photon pairs in many wavevector modes and stores the \emph{idler}
photon for controllable delay time. The \emph{idler} photon release
is controlled by real-time feedback allowing us to realize the ghost
imaging protocol without any additional image-preserving delay line, as comapred with Ref. \cite{Moreau2019}.
To the best of our knowledge, this is the first demonstration of quantum-memory-assisted GI experiment, and effectively a demonstration of atom-photon Bell-type entanglement in a single image.

Let us finally discuss potential applications and direct advantages over previosuly demonstrated ghost imaging experiments, based on spontaneous parametric down-conversion (SPDC). In the Table \ref{tab:comparison} we compare our results with other works that use delay line (DL) instead of quantum memory (QM) to acquire the ghost images. In the SPDC experiments the \emph{signal}-\emph{idler} coincidence rate (R) is limited by the maximum gating frequency of the image sensor and varies from few to few hundreds coincidences per second (cps). In our setup we achieve $\approx 6$ cps as we are limited by the atom cooling and trapping period that occupies 89\% of the sequence. The instantaneous rate is therefore around 60 cps and can be doubled in non-phase sensitive approach by removing the interferometers. This is comparable with most SPDC approaches although with faster camera gating SPDC-based GI could potentially reach much higher rates. The available number of modes (M) and achievable visibility (or equivalently signal to noise ratio) is comparable in both cases. The parameter that we want to focus on is the available time delay ($\tau$) needed to properly gate the camera which in case of SPDC experiments never exceeds $100\,$ns ($30\,$m of additional optical path). In our setup the delay can be set programatically and can be as long as the wavevector-dependent memory lifetime \cite{Lipka2021}, that for $45\, \mu$s allows retrieval of 50\% of the modes. Such long delay can be useful in applications as quantum ghost LIDAR that can be resistant to detection path turbulences \cite{Hardy2013} and quantum radar \cite{Shapiro2020,Pirandola2018} where the quantum illumination leads to improvement of signal to noise ratio \cite{Gregory2020}. Another difference is that the \emph{signal} and \emph{idler} photons emitted from our memory have few-MHz linewidths and thus they are easier to filter from the broadband background noise and can be used for remote sensing of inherently narrowband optomechanical and atomic systems, facilitating for example magnetic field imaging.
Moreover, presented system equipped with a more advanced \emph{signal} photon
detection and combined with wavefront shaping techniques based on
fast micro-mirror arrays or spatial light modulators could
lead to new adaptive quantum GI schemes \citep{Defienne2018}. As the phase modulation can be realized by fast all-optical
spin-wave modulation techniques \citep{Parniak2019,Mazelanik2019,Lipka2019}
that effectively reshape the \emph{idler }photon spatial structure,
the phase reshaping can be applied conditionally on \emph{signal} photon
measurement outcome. We also envisage that via feedback one may shape
the idler light not only spatially, but also in terms of its statistics
(akin to heralding single or multi-photon states) for protocols such
as quantum image distillation \citep{Defienne2019}. Our
technique could also be applied in quantum information schemes such
as quantum-secured imaging \citep{Malik2012}, quantum secure ghost
imaging \citep{Yao2018}.   
\begin{table}
\centering
\arrayrulecolor[rgb]{0.8,0.8,0.8}
\begin{tabular}{|c|c|c|l|c|c|c!{\color{black}\vrule}} 
\hline
Type & \begin{tabular}[c]{@{}c@{}}R\\$(\frac{1}{\mathrm{s}})$\end{tabular} & $\mathcal{V}$ & \multicolumn{1}{c|}{\begin{tabular}[c]{@{}c@{}}$\tau$\\($\mathrm{ns}$)\end{tabular}} & M   & PS  & \multicolumn{1}{c|}{Ref.}                                                \\ 
\hline
QM   & 6                                                                   & 0.78          & $>150$                                                                               & 370 & yes & \multicolumn{1}{c|}{\begin{tabular}[c]{@{}c@{}}This\\work\end{tabular}}  \\ 
\hline
DL   & 35                                                                  & 0.9           & 73                                                                                   & 500 & no  & \cite{Aspden2013}                                                        \\ 
\cline{1-6}\arrayrulecolor{black}\cline{7-7}
DL   & 528                                                                 & 0.71          & 84                                                                                   & 113 & no  & \cite{Aspden2015}                                                        \\ 
\arrayrulecolor[rgb]{0.8,0.8,0.8}\cline{1-6}\arrayrulecolor{black}\cline{7-7}
DL   & 4                                                                   & 0.86          & 67                                                                                   &     & yes & \cite{Moreau2019}                                                        \\ 
\arrayrulecolor[rgb]{0.8,0.8,0.8}\cline{1-6}\arrayrulecolor{black}\cline{7-7}
DL   & $<6$                                                                & 0.67          &                                                                                      &     & yes & \cite{Aspden2016}                                                        \\
\arrayrulecolor[rgb]{0.8,0.8,0.8}\cline{1-6}\arrayrulecolor{black}\cline{7-7}
\end{tabular}
\arrayrulecolor{black}
\caption{Comparison of two GI approaches: QM - Quantum Memory, DL - delay line in means of coincidence rate (R), visibility ($\mathcal{V}$), number of modes (M) and phase sensitivity (PS). \label{tab:comparison}}
\end{table}

\begin{acknowledgments}
We thank K. Banaszek for
the generous support and J. Kołodyński for insightful discussions. This work was funded by MNiSW (DI2016 014846, DI2018 010848),
National Science Centre (2016/21/B/ST2/02559, 2017/25/N/ST2/01163,
2017/25/N/ST2/00713) and by the Foundation for Polish Science (MAB/2018/4
``Quantum Optical Technologies''). The \textquotedbl Quantum Optical
Technologies ” project is carried out within the International Research
Agendas programme of the Foundation for Polish Science co-financed
by the European Union under the European Regional Development Fund.
M.P. was supported by the Foundation for Polish Science via the START
scholarship.
\end{acknowledgments}
\bibliographystyle{apsrev4-2}
\bibliography{gbell}

\begin{thebibliography}{51}%
\makeatletter
\providecommand \@ifxundefined [1]{%
 \@ifx{#1\undefined}
}%
\providecommand \@ifnum [1]{%
 \ifnum #1\expandafter \@firstoftwo
 \else \expandafter \@secondoftwo
 \fi
}%
\providecommand \@ifx [1]{%
 \ifx #1\expandafter \@firstoftwo
 \else \expandafter \@secondoftwo
 \fi
}%
\providecommand \natexlab [1]{#1}%
\providecommand \enquote  [1]{``#1''}%
\providecommand \bibnamefont  [1]{#1}%
\providecommand \bibfnamefont [1]{#1}%
\providecommand \citenamefont [1]{#1}%
\providecommand \href@noop [0]{\@secondoftwo}%
\providecommand \href [0]{\begingroup \@sanitize@url \@href}%
\providecommand \@href[1]{\@@startlink{#1}\@@href}%
\providecommand \@@href[1]{\endgroup#1\@@endlink}%
\providecommand \@sanitize@url [0]{\catcode `\\12\catcode `\$12\catcode
  `\&12\catcode `\#12\catcode `\^12\catcode `\_12\catcode `\%12\relax}%
\providecommand \@@startlink[1]{}%
\providecommand \@@endlink[0]{}%
\providecommand \url  [0]{\begingroup\@sanitize@url \@url }%
\providecommand \@url [1]{\endgroup\@href {#1}{\urlprefix }}%
\providecommand \urlprefix  [0]{URL }%
\providecommand \Eprint [0]{\href }%
\providecommand \doibase [0]{https://doi.org/}%
\providecommand \selectlanguage [0]{\@gobble}%
\providecommand \bibinfo  [0]{\@secondoftwo}%
\providecommand \bibfield  [0]{\@secondoftwo}%
\providecommand \translation [1]{[#1]}%
\providecommand \BibitemOpen [0]{}%
\providecommand \bibitemStop [0]{}%
\providecommand \bibitemNoStop [0]{.\EOS\space}%
\providecommand \EOS [0]{\spacefactor3000\relax}%
\providecommand \BibitemShut  [1]{\csname bibitem#1\endcsname}%
\let\auto@bib@innerbib\@empty
\bibitem [{\citenamefont {Freedman}\ and\ \citenamefont
  {Clauser}(1972)}]{Freedman1972}%
  \BibitemOpen
  \bibfield  {author} {\bibinfo {author} {\bibfnamefont {S.~J.}\ \bibnamefont
  {Freedman}}\ and\ \bibinfo {author} {\bibfnamefont {J.~F.}\ \bibnamefont
  {Clauser}},\ }\href {https://doi.org/10.1103/PhysRevLett.28.938} {\bibfield
  {journal} {\bibinfo  {journal} {Physical Review Letters}\ }\textbf {\bibinfo
  {volume} {28}},\ \bibinfo {pages} {938} (\bibinfo {year} {1972})}\BibitemShut
  {NoStop}%
\bibitem [{\citenamefont {Aspect}\ \emph {et~al.}(1982)\citenamefont {Aspect},
  \citenamefont {Grangier},\ and\ \citenamefont {Roger}}]{Aspect1982}%
  \BibitemOpen
  \bibfield  {author} {\bibinfo {author} {\bibfnamefont {A.}~\bibnamefont
  {Aspect}}, \bibinfo {author} {\bibfnamefont {P.}~\bibnamefont {Grangier}},\
  and\ \bibinfo {author} {\bibfnamefont {G.}~\bibnamefont {Roger}},\ }\href
  {https://doi.org/10.1103/PhysRevLett.49.91} {\bibfield  {journal} {\bibinfo
  {journal} {Physical Review Letters}\ }\textbf {\bibinfo {volume} {49}},\
  \bibinfo {pages} {91} (\bibinfo {year} {1982})}\BibitemShut {NoStop}%
\bibitem [{\citenamefont {Shalm}\ \emph {et~al.}(2015)\citenamefont {Shalm},
  \citenamefont {Meyer-Scott}, \citenamefont {Christensen}, \citenamefont
  {Bierhorst}, \citenamefont {Wayne}, \citenamefont {Stevens}, \citenamefont
  {Gerrits}, \citenamefont {Glancy}, \citenamefont {Hamel}, \citenamefont
  {Allman}, \citenamefont {Coakley}, \citenamefont {Dyer}, \citenamefont
  {Hodge}, \citenamefont {Lita}, \citenamefont {Verma}, \citenamefont
  {Lambrocco}, \citenamefont {Tortorici}, \citenamefont {Migdall},
  \citenamefont {Zhang}, \citenamefont {Kumor}, \citenamefont {Farr},
  \citenamefont {Marsili}, \citenamefont {Shaw}, \citenamefont {Stern},
  \citenamefont {Abell{\'{a}}n}, \citenamefont {Amaya}, \citenamefont
  {Pruneri}, \citenamefont {Jennewein}, \citenamefont {Mitchell}, \citenamefont
  {Kwiat}, \citenamefont {Bienfang}, \citenamefont {Mirin}, \citenamefont
  {Knill},\ and\ \citenamefont {Nam}}]{Shalm2015}%
  \BibitemOpen
  \bibfield  {author} {\bibinfo {author} {\bibfnamefont {L.~K.}\ \bibnamefont
  {Shalm}}, \bibinfo {author} {\bibfnamefont {E.}~\bibnamefont {Meyer-Scott}},
  \bibinfo {author} {\bibfnamefont {B.~G.}\ \bibnamefont {Christensen}},
  \bibinfo {author} {\bibfnamefont {P.}~\bibnamefont {Bierhorst}}, \bibinfo
  {author} {\bibfnamefont {M.~A.}\ \bibnamefont {Wayne}}, \bibinfo {author}
  {\bibfnamefont {M.~J.}\ \bibnamefont {Stevens}}, \bibinfo {author}
  {\bibfnamefont {T.}~\bibnamefont {Gerrits}}, \bibinfo {author} {\bibfnamefont
  {S.}~\bibnamefont {Glancy}}, \bibinfo {author} {\bibfnamefont {D.~R.}\
  \bibnamefont {Hamel}}, \bibinfo {author} {\bibfnamefont {M.~S.}\ \bibnamefont
  {Allman}}, \bibinfo {author} {\bibfnamefont {K.~J.}\ \bibnamefont {Coakley}},
  \bibinfo {author} {\bibfnamefont {S.~D.}\ \bibnamefont {Dyer}}, \bibinfo
  {author} {\bibfnamefont {C.}~\bibnamefont {Hodge}}, \bibinfo {author}
  {\bibfnamefont {A.~E.}\ \bibnamefont {Lita}}, \bibinfo {author}
  {\bibfnamefont {V.~B.}\ \bibnamefont {Verma}}, \bibinfo {author}
  {\bibfnamefont {C.}~\bibnamefont {Lambrocco}}, \bibinfo {author}
  {\bibfnamefont {E.}~\bibnamefont {Tortorici}}, \bibinfo {author}
  {\bibfnamefont {A.~L.}\ \bibnamefont {Migdall}}, \bibinfo {author}
  {\bibfnamefont {Y.}~\bibnamefont {Zhang}}, \bibinfo {author} {\bibfnamefont
  {D.~R.}\ \bibnamefont {Kumor}}, \bibinfo {author} {\bibfnamefont {W.~H.}\
  \bibnamefont {Farr}}, \bibinfo {author} {\bibfnamefont {F.}~\bibnamefont
  {Marsili}}, \bibinfo {author} {\bibfnamefont {M.~D.}\ \bibnamefont {Shaw}},
  \bibinfo {author} {\bibfnamefont {J.~A.}\ \bibnamefont {Stern}}, \bibinfo
  {author} {\bibfnamefont {C.}~\bibnamefont {Abell{\'{a}}n}}, \bibinfo {author}
  {\bibfnamefont {W.}~\bibnamefont {Amaya}}, \bibinfo {author} {\bibfnamefont
  {V.}~\bibnamefont {Pruneri}}, \bibinfo {author} {\bibfnamefont
  {T.}~\bibnamefont {Jennewein}}, \bibinfo {author} {\bibfnamefont {M.~W.}\
  \bibnamefont {Mitchell}}, \bibinfo {author} {\bibfnamefont {P.~G.}\
  \bibnamefont {Kwiat}}, \bibinfo {author} {\bibfnamefont {J.~C.}\ \bibnamefont
  {Bienfang}}, \bibinfo {author} {\bibfnamefont {R.~P.}\ \bibnamefont {Mirin}},
  \bibinfo {author} {\bibfnamefont {E.}~\bibnamefont {Knill}},\ and\ \bibinfo
  {author} {\bibfnamefont {S.~W.}\ \bibnamefont {Nam}},\ }\href
  {https://doi.org/10.1103/PhysRevLett.115.250402} {\bibfield  {journal}
  {\bibinfo  {journal} {Physical Review Letters}\ }\textbf {\bibinfo {volume}
  {115}},\ \bibinfo {pages} {250402} (\bibinfo {year} {2015})}\BibitemShut
  {NoStop}%
\bibitem [{\citenamefont {Giustina}\ \emph {et~al.}(2015)\citenamefont
  {Giustina}, \citenamefont {Versteegh}, \citenamefont {Wengerowsky},
  \citenamefont {Handsteiner}, \citenamefont {Hochrainer}, \citenamefont
  {Phelan}, \citenamefont {Steinlechner}, \citenamefont {Kofler}, \citenamefont
  {Larsson}, \citenamefont {Abell{\'{a}}n}, \citenamefont {Amaya},
  \citenamefont {Pruneri}, \citenamefont {Mitchell}, \citenamefont {Beyer},
  \citenamefont {Gerrits}, \citenamefont {Lita}, \citenamefont {Shalm},
  \citenamefont {Nam}, \citenamefont {Scheidl}, \citenamefont {Ursin},
  \citenamefont {Wittmann},\ and\ \citenamefont {Zeilinger}}]{Giustina2015}%
  \BibitemOpen
  \bibfield  {author} {\bibinfo {author} {\bibfnamefont {M.}~\bibnamefont
  {Giustina}}, \bibinfo {author} {\bibfnamefont {M.~A.}\ \bibnamefont
  {Versteegh}}, \bibinfo {author} {\bibfnamefont {S.}~\bibnamefont
  {Wengerowsky}}, \bibinfo {author} {\bibfnamefont {J.}~\bibnamefont
  {Handsteiner}}, \bibinfo {author} {\bibfnamefont {A.}~\bibnamefont
  {Hochrainer}}, \bibinfo {author} {\bibfnamefont {K.}~\bibnamefont {Phelan}},
  \bibinfo {author} {\bibfnamefont {F.}~\bibnamefont {Steinlechner}}, \bibinfo
  {author} {\bibfnamefont {J.}~\bibnamefont {Kofler}}, \bibinfo {author}
  {\bibfnamefont {J.~{\AA}.}\ \bibnamefont {Larsson}}, \bibinfo {author}
  {\bibfnamefont {C.}~\bibnamefont {Abell{\'{a}}n}}, \bibinfo {author}
  {\bibfnamefont {W.}~\bibnamefont {Amaya}}, \bibinfo {author} {\bibfnamefont
  {V.}~\bibnamefont {Pruneri}}, \bibinfo {author} {\bibfnamefont {M.~W.}\
  \bibnamefont {Mitchell}}, \bibinfo {author} {\bibfnamefont {J.}~\bibnamefont
  {Beyer}}, \bibinfo {author} {\bibfnamefont {T.}~\bibnamefont {Gerrits}},
  \bibinfo {author} {\bibfnamefont {A.~E.}\ \bibnamefont {Lita}}, \bibinfo
  {author} {\bibfnamefont {L.~K.}\ \bibnamefont {Shalm}}, \bibinfo {author}
  {\bibfnamefont {S.~W.}\ \bibnamefont {Nam}}, \bibinfo {author} {\bibfnamefont
  {T.}~\bibnamefont {Scheidl}}, \bibinfo {author} {\bibfnamefont
  {R.}~\bibnamefont {Ursin}}, \bibinfo {author} {\bibfnamefont
  {B.}~\bibnamefont {Wittmann}},\ and\ \bibinfo {author} {\bibfnamefont
  {A.}~\bibnamefont {Zeilinger}},\ }\href
  {https://doi.org/10.1103/PhysRevLett.115.250401} {\bibfield  {journal}
  {\bibinfo  {journal} {Physical Review Letters}\ }\textbf {\bibinfo {volume}
  {115}},\ \bibinfo {pages} {250401} (\bibinfo {year} {2015})}\BibitemShut
  {NoStop}%
\bibitem [{\citenamefont {Abell{\'{a}}n}\ \emph {et~al.}(2018)\citenamefont
  {Abell{\'{a}}n}, \citenamefont {Ac{\'{i}}n}, \citenamefont {Alarc{\'{o}}n},
  \citenamefont {Alibart}, \citenamefont {Andersen}, \citenamefont {Andreoli},
  \citenamefont {Beckert}, \citenamefont {Beduini}, \citenamefont {Bendersky},
  \citenamefont {Bentivegna}, \citenamefont {Bierhorst}, \citenamefont
  {Burchardt}, \citenamefont {Cabello}, \citenamefont {Cari{\~{n}}e},
  \citenamefont {Carrasco}, \citenamefont {Carvacho}, \citenamefont
  {Cavalcanti}, \citenamefont {Chaves}, \citenamefont {Cort{\'{e}}s-Vega},
  \citenamefont {Cuevas}, \citenamefont {Delgado}, \citenamefont {{De
  Riedmatten}}, \citenamefont {Eichler}, \citenamefont {Farrera}, \citenamefont
  {Fuenzalida}, \citenamefont {Garc{\'{i}}a-Matos}, \citenamefont {Garthoff},
  \citenamefont {Gasparinetti}, \citenamefont {Gerrits}, \citenamefont
  {{Ghafari Jouneghani}}, \citenamefont {Glancy}, \citenamefont {G{\'{o}}mez},
  \citenamefont {Gonz{\'{a}}lez}, \citenamefont {Guan}, \citenamefont
  {Handsteiner}, \citenamefont {Heinsoo}, \citenamefont {Heinze}, \citenamefont
  {Hirschmann}, \citenamefont {Jim{\'{e}}nez}, \citenamefont {Kaiser},
  \citenamefont {Knill}, \citenamefont {Knoll}, \citenamefont {Krinner},
  \citenamefont {Kurpiers}, \citenamefont {Larotonda}, \citenamefont {Larsson},
  \citenamefont {Lenhard}, \citenamefont {Li}, \citenamefont {Li},
  \citenamefont {Lima}, \citenamefont {Liu}, \citenamefont {Liu}, \citenamefont
  {Grande}, \citenamefont {Lunghi}, \citenamefont {Ma}, \citenamefont
  {Maga{\~{n}}a-Loaiza}, \citenamefont {Magnard}, \citenamefont {Magnoni},
  \citenamefont {Mart{\'{i}}-Prieto}, \citenamefont {Mart{\'{i}}nez},
  \citenamefont {Mataloni}, \citenamefont {Mattar}, \citenamefont {Mazzera},
  \citenamefont {Mirin}, \citenamefont {Mitchell}, \citenamefont {Nam},
  \citenamefont {Oppliger}, \citenamefont {Pan}, \citenamefont {Patel},
  \citenamefont {Pryde}, \citenamefont {Rauch}, \citenamefont {Redeker},
  \citenamefont {Riel{\"{a}}nder}, \citenamefont {Ringbauer}, \citenamefont
  {Roberson}, \citenamefont {Rosenfeld}, \citenamefont {Salath{\'{e}}},
  \citenamefont {Santodonato}, \citenamefont {Sauder}, \citenamefont {Scheidl},
  \citenamefont {Schmiegelow}, \citenamefont {Sciarrino}, \citenamefont {Seri},
  \citenamefont {Shalm}, \citenamefont {Shi}, \citenamefont {Slussarenko},
  \citenamefont {Stevens}, \citenamefont {Tanzilli}, \citenamefont {Toledo},
  \citenamefont {Tura}, \citenamefont {Ursin}, \citenamefont {Vergyris},
  \citenamefont {Verma}, \citenamefont {Walter}, \citenamefont {Wallraff},
  \citenamefont {Wang}, \citenamefont {Weinfurter}, \citenamefont {Weston},
  \citenamefont {White}, \citenamefont {Wu}, \citenamefont {Xavier},
  \citenamefont {You}, \citenamefont {Yuan}, \citenamefont {Zeilinger},
  \citenamefont {Zhang}, \citenamefont {Zhang},\ and\ \citenamefont
  {Zhong}}]{Abellan2018}%
  \BibitemOpen
  \bibfield  {author} {\bibinfo {author} {\bibfnamefont {C.}~\bibnamefont
  {Abell{\'{a}}n}}, \bibinfo {author} {\bibfnamefont {A.}~\bibnamefont
  {Ac{\'{i}}n}}, \bibinfo {author} {\bibfnamefont {A.}~\bibnamefont
  {Alarc{\'{o}}n}}, \bibinfo {author} {\bibfnamefont {O.}~\bibnamefont
  {Alibart}}, \bibinfo {author} {\bibfnamefont {C.~K.}\ \bibnamefont
  {Andersen}}, \bibinfo {author} {\bibfnamefont {F.}~\bibnamefont {Andreoli}},
  \bibinfo {author} {\bibfnamefont {A.}~\bibnamefont {Beckert}}, \bibinfo
  {author} {\bibfnamefont {F.~A.}\ \bibnamefont {Beduini}}, \bibinfo {author}
  {\bibfnamefont {A.}~\bibnamefont {Bendersky}}, \bibinfo {author}
  {\bibfnamefont {M.}~\bibnamefont {Bentivegna}}, \bibinfo {author}
  {\bibfnamefont {P.}~\bibnamefont {Bierhorst}}, \bibinfo {author}
  {\bibfnamefont {D.}~\bibnamefont {Burchardt}}, \bibinfo {author}
  {\bibfnamefont {A.}~\bibnamefont {Cabello}}, \bibinfo {author} {\bibfnamefont
  {J.}~\bibnamefont {Cari{\~{n}}e}}, \bibinfo {author} {\bibfnamefont
  {S.}~\bibnamefont {Carrasco}}, \bibinfo {author} {\bibfnamefont
  {G.}~\bibnamefont {Carvacho}}, \bibinfo {author} {\bibfnamefont
  {D.}~\bibnamefont {Cavalcanti}}, \bibinfo {author} {\bibfnamefont
  {R.}~\bibnamefont {Chaves}}, \bibinfo {author} {\bibfnamefont
  {J.}~\bibnamefont {Cort{\'{e}}s-Vega}}, \bibinfo {author} {\bibfnamefont
  {A.}~\bibnamefont {Cuevas}}, \bibinfo {author} {\bibfnamefont
  {A.}~\bibnamefont {Delgado}}, \bibinfo {author} {\bibfnamefont
  {H.}~\bibnamefont {{De Riedmatten}}}, \bibinfo {author} {\bibfnamefont
  {C.}~\bibnamefont {Eichler}}, \bibinfo {author} {\bibfnamefont
  {P.}~\bibnamefont {Farrera}}, \bibinfo {author} {\bibfnamefont
  {J.}~\bibnamefont {Fuenzalida}}, \bibinfo {author} {\bibfnamefont
  {M.}~\bibnamefont {Garc{\'{i}}a-Matos}}, \bibinfo {author} {\bibfnamefont
  {R.}~\bibnamefont {Garthoff}}, \bibinfo {author} {\bibfnamefont
  {S.}~\bibnamefont {Gasparinetti}}, \bibinfo {author} {\bibfnamefont
  {T.}~\bibnamefont {Gerrits}}, \bibinfo {author} {\bibfnamefont
  {F.}~\bibnamefont {{Ghafari Jouneghani}}}, \bibinfo {author} {\bibfnamefont
  {S.}~\bibnamefont {Glancy}}, \bibinfo {author} {\bibfnamefont {E.~S.}\
  \bibnamefont {G{\'{o}}mez}}, \bibinfo {author} {\bibfnamefont
  {P.}~\bibnamefont {Gonz{\'{a}}lez}}, \bibinfo {author} {\bibfnamefont
  {J.~Y.}\ \bibnamefont {Guan}}, \bibinfo {author} {\bibfnamefont
  {J.}~\bibnamefont {Handsteiner}}, \bibinfo {author} {\bibfnamefont
  {J.}~\bibnamefont {Heinsoo}}, \bibinfo {author} {\bibfnamefont
  {G.}~\bibnamefont {Heinze}}, \bibinfo {author} {\bibfnamefont
  {A.}~\bibnamefont {Hirschmann}}, \bibinfo {author} {\bibfnamefont
  {O.}~\bibnamefont {Jim{\'{e}}nez}}, \bibinfo {author} {\bibfnamefont
  {F.}~\bibnamefont {Kaiser}}, \bibinfo {author} {\bibfnamefont
  {E.}~\bibnamefont {Knill}}, \bibinfo {author} {\bibfnamefont {L.~T.}\
  \bibnamefont {Knoll}}, \bibinfo {author} {\bibfnamefont {S.}~\bibnamefont
  {Krinner}}, \bibinfo {author} {\bibfnamefont {P.}~\bibnamefont {Kurpiers}},
  \bibinfo {author} {\bibfnamefont {M.~A.}\ \bibnamefont {Larotonda}}, \bibinfo
  {author} {\bibfnamefont {J.~A.}\ \bibnamefont {Larsson}}, \bibinfo {author}
  {\bibfnamefont {A.}~\bibnamefont {Lenhard}}, \bibinfo {author} {\bibfnamefont
  {H.}~\bibnamefont {Li}}, \bibinfo {author} {\bibfnamefont {M.~H.}\
  \bibnamefont {Li}}, \bibinfo {author} {\bibfnamefont {G.}~\bibnamefont
  {Lima}}, \bibinfo {author} {\bibfnamefont {B.}~\bibnamefont {Liu}}, \bibinfo
  {author} {\bibfnamefont {Y.}~\bibnamefont {Liu}}, \bibinfo {author}
  {\bibfnamefont {I.~H.}\ \bibnamefont {Grande}}, \bibinfo {author}
  {\bibfnamefont {T.}~\bibnamefont {Lunghi}}, \bibinfo {author} {\bibfnamefont
  {X.}~\bibnamefont {Ma}}, \bibinfo {author} {\bibfnamefont {O.~S.}\
  \bibnamefont {Maga{\~{n}}a-Loaiza}}, \bibinfo {author} {\bibfnamefont
  {P.}~\bibnamefont {Magnard}}, \bibinfo {author} {\bibfnamefont
  {A.}~\bibnamefont {Magnoni}}, \bibinfo {author} {\bibfnamefont
  {M.}~\bibnamefont {Mart{\'{i}}-Prieto}}, \bibinfo {author} {\bibfnamefont
  {D.}~\bibnamefont {Mart{\'{i}}nez}}, \bibinfo {author} {\bibfnamefont
  {P.}~\bibnamefont {Mataloni}}, \bibinfo {author} {\bibfnamefont
  {A.}~\bibnamefont {Mattar}}, \bibinfo {author} {\bibfnamefont
  {M.}~\bibnamefont {Mazzera}}, \bibinfo {author} {\bibfnamefont {R.~P.}\
  \bibnamefont {Mirin}}, \bibinfo {author} {\bibfnamefont {M.~W.}\ \bibnamefont
  {Mitchell}}, \bibinfo {author} {\bibfnamefont {S.}~\bibnamefont {Nam}},
  \bibinfo {author} {\bibfnamefont {M.}~\bibnamefont {Oppliger}}, \bibinfo
  {author} {\bibfnamefont {J.~W.}\ \bibnamefont {Pan}}, \bibinfo {author}
  {\bibfnamefont {R.~B.}\ \bibnamefont {Patel}}, \bibinfo {author}
  {\bibfnamefont {G.~J.}\ \bibnamefont {Pryde}}, \bibinfo {author}
  {\bibfnamefont {D.}~\bibnamefont {Rauch}}, \bibinfo {author} {\bibfnamefont
  {K.}~\bibnamefont {Redeker}}, \bibinfo {author} {\bibfnamefont
  {D.}~\bibnamefont {Riel{\"{a}}nder}}, \bibinfo {author} {\bibfnamefont
  {M.}~\bibnamefont {Ringbauer}}, \bibinfo {author} {\bibfnamefont
  {T.}~\bibnamefont {Roberson}}, \bibinfo {author} {\bibfnamefont
  {W.}~\bibnamefont {Rosenfeld}}, \bibinfo {author} {\bibfnamefont
  {Y.}~\bibnamefont {Salath{\'{e}}}}, \bibinfo {author} {\bibfnamefont
  {L.}~\bibnamefont {Santodonato}}, \bibinfo {author} {\bibfnamefont
  {G.}~\bibnamefont {Sauder}}, \bibinfo {author} {\bibfnamefont
  {T.}~\bibnamefont {Scheidl}}, \bibinfo {author} {\bibfnamefont {C.~T.}\
  \bibnamefont {Schmiegelow}}, \bibinfo {author} {\bibfnamefont
  {F.}~\bibnamefont {Sciarrino}}, \bibinfo {author} {\bibfnamefont
  {A.}~\bibnamefont {Seri}}, \bibinfo {author} {\bibfnamefont {L.~K.}\
  \bibnamefont {Shalm}}, \bibinfo {author} {\bibfnamefont {S.~C.}\ \bibnamefont
  {Shi}}, \bibinfo {author} {\bibfnamefont {S.}~\bibnamefont {Slussarenko}},
  \bibinfo {author} {\bibfnamefont {M.~J.}\ \bibnamefont {Stevens}}, \bibinfo
  {author} {\bibfnamefont {S.}~\bibnamefont {Tanzilli}}, \bibinfo {author}
  {\bibfnamefont {F.}~\bibnamefont {Toledo}}, \bibinfo {author} {\bibfnamefont
  {J.}~\bibnamefont {Tura}}, \bibinfo {author} {\bibfnamefont {R.}~\bibnamefont
  {Ursin}}, \bibinfo {author} {\bibfnamefont {P.}~\bibnamefont {Vergyris}},
  \bibinfo {author} {\bibfnamefont {V.~B.}\ \bibnamefont {Verma}}, \bibinfo
  {author} {\bibfnamefont {T.}~\bibnamefont {Walter}}, \bibinfo {author}
  {\bibfnamefont {A.}~\bibnamefont {Wallraff}}, \bibinfo {author}
  {\bibfnamefont {Z.}~\bibnamefont {Wang}}, \bibinfo {author} {\bibfnamefont
  {H.}~\bibnamefont {Weinfurter}}, \bibinfo {author} {\bibfnamefont {M.~M.}\
  \bibnamefont {Weston}}, \bibinfo {author} {\bibfnamefont {A.~G.}\
  \bibnamefont {White}}, \bibinfo {author} {\bibfnamefont {C.}~\bibnamefont
  {Wu}}, \bibinfo {author} {\bibfnamefont {G.~B.}\ \bibnamefont {Xavier}},
  \bibinfo {author} {\bibfnamefont {L.}~\bibnamefont {You}}, \bibinfo {author}
  {\bibfnamefont {X.}~\bibnamefont {Yuan}}, \bibinfo {author} {\bibfnamefont
  {A.}~\bibnamefont {Zeilinger}}, \bibinfo {author} {\bibfnamefont
  {Q.}~\bibnamefont {Zhang}}, \bibinfo {author} {\bibfnamefont
  {W.}~\bibnamefont {Zhang}},\ and\ \bibinfo {author} {\bibfnamefont
  {J.}~\bibnamefont {Zhong}},\ }\href
  {https://doi.org/10.1038/s41586-018-0085-3} {\bibfield  {journal} {\bibinfo
  {journal} {Nature}\ }\textbf {\bibinfo {volume} {557}},\ \bibinfo {pages}
  {212} (\bibinfo {year} {2018})}\BibitemShut {NoStop}%
\bibitem [{\citenamefont {Vedovato}\ \emph {et~al.}(2018)\citenamefont
  {Vedovato}, \citenamefont {Agnesi}, \citenamefont {Tomasin}, \citenamefont
  {Avesani}, \citenamefont {Larsson}, \citenamefont {Vallone},\ and\
  \citenamefont {Villoresi}}]{Vedovato2018}%
  \BibitemOpen
  \bibfield  {author} {\bibinfo {author} {\bibfnamefont {F.}~\bibnamefont
  {Vedovato}}, \bibinfo {author} {\bibfnamefont {C.}~\bibnamefont {Agnesi}},
  \bibinfo {author} {\bibfnamefont {M.}~\bibnamefont {Tomasin}}, \bibinfo
  {author} {\bibfnamefont {M.}~\bibnamefont {Avesani}}, \bibinfo {author}
  {\bibfnamefont {J.~{\AA}.}\ \bibnamefont {Larsson}}, \bibinfo {author}
  {\bibfnamefont {G.}~\bibnamefont {Vallone}},\ and\ \bibinfo {author}
  {\bibfnamefont {P.}~\bibnamefont {Villoresi}},\ }\href
  {https://doi.org/10.1103/PhysRevLett.121.190401} {\bibfield  {journal}
  {\bibinfo  {journal} {Physical Review Letters}\ }\textbf {\bibinfo {volume}
  {121}},\ \bibinfo {pages} {190401} (\bibinfo {year} {2018})}\BibitemShut
  {NoStop}%
\bibitem [{\citenamefont {Brendel}\ \emph {et~al.}(1999)\citenamefont
  {Brendel}, \citenamefont {Gisin}, \citenamefont {Tittel},\ and\ \citenamefont
  {Zbinden}}]{Brendel1999}%
  \BibitemOpen
  \bibfield  {author} {\bibinfo {author} {\bibfnamefont {J.}~\bibnamefont
  {Brendel}}, \bibinfo {author} {\bibfnamefont {N.}~\bibnamefont {Gisin}},
  \bibinfo {author} {\bibfnamefont {W.}~\bibnamefont {Tittel}},\ and\ \bibinfo
  {author} {\bibfnamefont {H.}~\bibnamefont {Zbinden}},\ }\href
  {https://doi.org/10.1103/PhysRevLett.82.2594} {\bibfield  {journal} {\bibinfo
   {journal} {Physical Review Letters}\ }\textbf {\bibinfo {volume} {82}},\
  \bibinfo {pages} {2594} (\bibinfo {year} {1999})}\BibitemShut {NoStop}%
\bibitem [{\citenamefont {Yarnall}\ \emph {et~al.}(2007)\citenamefont
  {Yarnall}, \citenamefont {Abouraddy}, \citenamefont {Saleh},\ and\
  \citenamefont {Teich}}]{Yarnall2007}%
  \BibitemOpen
  \bibfield  {author} {\bibinfo {author} {\bibfnamefont {T.}~\bibnamefont
  {Yarnall}}, \bibinfo {author} {\bibfnamefont {A.~F.}\ \bibnamefont
  {Abouraddy}}, \bibinfo {author} {\bibfnamefont {B.~E.}\ \bibnamefont
  {Saleh}},\ and\ \bibinfo {author} {\bibfnamefont {M.~C.}\ \bibnamefont
  {Teich}},\ }\href {https://doi.org/10.1103/PhysRevLett.99.170408} {\bibfield
  {journal} {\bibinfo  {journal} {Physical Review Letters}\ }\textbf {\bibinfo
  {volume} {99}},\ \bibinfo {pages} {170408} (\bibinfo {year}
  {2007})}\BibitemShut {NoStop}%
\bibitem [{\citenamefont {Rarity}\ and\ \citenamefont
  {Tapster}(1990)}]{Rarity1990}%
  \BibitemOpen
  \bibfield  {author} {\bibinfo {author} {\bibfnamefont {J.~G.}\ \bibnamefont
  {Rarity}}\ and\ \bibinfo {author} {\bibfnamefont {P.~R.}\ \bibnamefont
  {Tapster}},\ }\href {https://doi.org/10.1103/PhysRevLett.64.2495} {\bibfield
  {journal} {\bibinfo  {journal} {Physical Review Letters}\ }\textbf {\bibinfo
  {volume} {64}},\ \bibinfo {pages} {2495} (\bibinfo {year}
  {1990})}\BibitemShut {NoStop}%
\bibitem [{\citenamefont {Leach}\ \emph {et~al.}(2009)\citenamefont {Leach},
  \citenamefont {Jack}, \citenamefont {Romero}, \citenamefont {Ritsch-Marte},
  \citenamefont {Boyd}, \citenamefont {Jha}, \citenamefont {Barnett},
  \citenamefont {Franke-Arnold},\ and\ \citenamefont {Padgett}}]{Leach2009}%
  \BibitemOpen
  \bibfield  {author} {\bibinfo {author} {\bibfnamefont {J.}~\bibnamefont
  {Leach}}, \bibinfo {author} {\bibfnamefont {B.}~\bibnamefont {Jack}},
  \bibinfo {author} {\bibfnamefont {J.}~\bibnamefont {Romero}}, \bibinfo
  {author} {\bibfnamefont {M.}~\bibnamefont {Ritsch-Marte}}, \bibinfo {author}
  {\bibfnamefont {R.~W.}\ \bibnamefont {Boyd}}, \bibinfo {author}
  {\bibfnamefont {A.~K.}\ \bibnamefont {Jha}}, \bibinfo {author} {\bibfnamefont
  {S.~M.}\ \bibnamefont {Barnett}}, \bibinfo {author} {\bibfnamefont
  {S.}~\bibnamefont {Franke-Arnold}},\ and\ \bibinfo {author} {\bibfnamefont
  {M.~J.}\ \bibnamefont {Padgett}},\ }\href
  {https://doi.org/10.1364/oe.17.008287} {\bibfield  {journal} {\bibinfo
  {journal} {Optics Express}\ }\textbf {\bibinfo {volume} {17}},\ \bibinfo
  {pages} {8287} (\bibinfo {year} {2009})}\BibitemShut {NoStop}%
\bibitem [{\citenamefont {Bell}(1964)}]{Bell1964}%
  \BibitemOpen
  \bibfield  {author} {\bibinfo {author} {\bibfnamefont {J.~S.}\ \bibnamefont
  {Bell}},\ }\href {https://doi.org/10.1103/physicsphysiquefizika.1.195}
  {\bibfield  {journal} {\bibinfo  {journal} {Physics}\ }\textbf {\bibinfo
  {volume} {1}},\ \bibinfo {pages} {195} (\bibinfo {year} {1964})}\BibitemShut
  {NoStop}%
\bibitem [{\citenamefont {Ac{\'{i}}n}\ \emph {et~al.}(2006)\citenamefont
  {Ac{\'{i}}n}, \citenamefont {Gisin},\ and\ \citenamefont
  {Masanes}}]{Acin2006}%
  \BibitemOpen
  \bibfield  {author} {\bibinfo {author} {\bibfnamefont {A.}~\bibnamefont
  {Ac{\'{i}}n}}, \bibinfo {author} {\bibfnamefont {N.}~\bibnamefont {Gisin}},\
  and\ \bibinfo {author} {\bibfnamefont {L.}~\bibnamefont {Masanes}},\ }\href
  {https://doi.org/10.1103/PhysRevLett.97.120405} {\bibfield  {journal}
  {\bibinfo  {journal} {Physical Review Letters}\ }\textbf {\bibinfo {volume}
  {97}},\ \bibinfo {pages} {120405} (\bibinfo {year} {2006})}\BibitemShut
  {NoStop}%
\bibitem [{\citenamefont {Yuan}\ \emph {et~al.}(2008)\citenamefont {Yuan},
  \citenamefont {Chen}, \citenamefont {Zhao}, \citenamefont {Chen},
  \citenamefont {Schmiedmayer},\ and\ \citenamefont {Pan}}]{Yuan2008}%
  \BibitemOpen
  \bibfield  {author} {\bibinfo {author} {\bibfnamefont {Z.~S.}\ \bibnamefont
  {Yuan}}, \bibinfo {author} {\bibfnamefont {Y.~A.}\ \bibnamefont {Chen}},
  \bibinfo {author} {\bibfnamefont {B.}~\bibnamefont {Zhao}}, \bibinfo {author}
  {\bibfnamefont {S.}~\bibnamefont {Chen}}, \bibinfo {author} {\bibfnamefont
  {J.}~\bibnamefont {Schmiedmayer}},\ and\ \bibinfo {author} {\bibfnamefont
  {J.~W.}\ \bibnamefont {Pan}},\ }\href {https://doi.org/10.1038/nature07241}
  {\bibfield  {journal} {\bibinfo  {journal} {Nature}\ }\textbf {\bibinfo
  {volume} {454}},\ \bibinfo {pages} {1098} (\bibinfo {year}
  {2008})}\BibitemShut {NoStop}%
\bibitem [{\citenamefont {Einstein}\ \emph {et~al.}(1935)\citenamefont
  {Einstein}, \citenamefont {Podolsky},\ and\ \citenamefont
  {Rosen}}]{Einstein1935}%
  \BibitemOpen
  \bibfield  {author} {\bibinfo {author} {\bibfnamefont {A.}~\bibnamefont
  {Einstein}}, \bibinfo {author} {\bibfnamefont {B.}~\bibnamefont {Podolsky}},\
  and\ \bibinfo {author} {\bibfnamefont {N.}~\bibnamefont {Rosen}},\ }\href
  {https://doi.org/10.1103/PhysRev.47.777} {\bibfield  {journal} {\bibinfo
  {journal} {Physical Review}\ }\textbf {\bibinfo {volume} {47}},\ \bibinfo
  {pages} {777} (\bibinfo {year} {1935})}\BibitemShut {NoStop}%
\bibitem [{\citenamefont {Jensen}\ \emph {et~al.}(2011)\citenamefont {Jensen},
  \citenamefont {Wasilewski}, \citenamefont {Krauter}, \citenamefont
  {Fernholz}, \citenamefont {Nielsen}, \citenamefont {Owari}, \citenamefont
  {Plenio}, \citenamefont {Serafini}, \citenamefont {Wolf},\ and\ \citenamefont
  {Polzik}}]{Jensen2011}%
  \BibitemOpen
  \bibfield  {author} {\bibinfo {author} {\bibfnamefont {K.}~\bibnamefont
  {Jensen}}, \bibinfo {author} {\bibfnamefont {W.}~\bibnamefont {Wasilewski}},
  \bibinfo {author} {\bibfnamefont {H.}~\bibnamefont {Krauter}}, \bibinfo
  {author} {\bibfnamefont {T.}~\bibnamefont {Fernholz}}, \bibinfo {author}
  {\bibfnamefont {B.~M.}\ \bibnamefont {Nielsen}}, \bibinfo {author}
  {\bibfnamefont {M.}~\bibnamefont {Owari}}, \bibinfo {author} {\bibfnamefont
  {M.~B.}\ \bibnamefont {Plenio}}, \bibinfo {author} {\bibfnamefont
  {A.}~\bibnamefont {Serafini}}, \bibinfo {author} {\bibfnamefont {M.~M.}\
  \bibnamefont {Wolf}},\ and\ \bibinfo {author} {\bibfnamefont {E.~S.}\
  \bibnamefont {Polzik}},\ }\href {https://doi.org/10.1038/nphys1819}
  {\bibfield  {journal} {\bibinfo  {journal} {Nature Physics}\ }\textbf
  {\bibinfo {volume} {7}},\ \bibinfo {pages} {13} (\bibinfo {year}
  {2011})}\BibitemShut {NoStop}%
\bibitem [{\citenamefont {Ou}\ \emph {et~al.}(1992)\citenamefont {Ou},
  \citenamefont {Pereira}, \citenamefont {Kimble},\ and\ \citenamefont
  {Peng}}]{Ou1992}%
  \BibitemOpen
  \bibfield  {author} {\bibinfo {author} {\bibfnamefont {Z.~Y.}\ \bibnamefont
  {Ou}}, \bibinfo {author} {\bibfnamefont {S.~F.}\ \bibnamefont {Pereira}},
  \bibinfo {author} {\bibfnamefont {H.~J.}\ \bibnamefont {Kimble}},\ and\
  \bibinfo {author} {\bibfnamefont {K.~C.}\ \bibnamefont {Peng}},\ }\href
  {https://doi.org/10.1103/PhysRevLett.68.3663} {\bibfield  {journal} {\bibinfo
   {journal} {Physical Review Letters}\ }\textbf {\bibinfo {volume} {68}},\
  \bibinfo {pages} {3663} (\bibinfo {year} {1992})}\BibitemShut {NoStop}%
\bibitem [{\citenamefont {Howell}\ \emph {et~al.}(2004)\citenamefont {Howell},
  \citenamefont {Bennink}, \citenamefont {Bentley},\ and\ \citenamefont
  {Boyd}}]{Howell2004}%
  \BibitemOpen
  \bibfield  {author} {\bibinfo {author} {\bibfnamefont {J.~C.}\ \bibnamefont
  {Howell}}, \bibinfo {author} {\bibfnamefont {R.~S.}\ \bibnamefont {Bennink}},
  \bibinfo {author} {\bibfnamefont {S.~J.}\ \bibnamefont {Bentley}},\ and\
  \bibinfo {author} {\bibfnamefont {R.~W.}\ \bibnamefont {Boyd}},\ }\href
  {https://doi.org/10.1103/PhysRevLett.92.210403} {\bibfield  {journal}
  {\bibinfo  {journal} {Physical Review Letters}\ }\textbf {\bibinfo {volume}
  {92}},\ \bibinfo {pages} {210403} (\bibinfo {year} {2004})}\BibitemShut
  {NoStop}%
\bibitem [{\citenamefont {Edgar}\ \emph {et~al.}(2012)\citenamefont {Edgar},
  \citenamefont {Tasca}, \citenamefont {Izdebski}, \citenamefont {Warburton},
  \citenamefont {Leach}, \citenamefont {Agnew}, \citenamefont {Buller},
  \citenamefont {Boyd},\ and\ \citenamefont {Padgett}}]{Edgar2012}%
  \BibitemOpen
  \bibfield  {author} {\bibinfo {author} {\bibfnamefont {M.~P.}\ \bibnamefont
  {Edgar}}, \bibinfo {author} {\bibfnamefont {D.~S.}\ \bibnamefont {Tasca}},
  \bibinfo {author} {\bibfnamefont {F.}~\bibnamefont {Izdebski}}, \bibinfo
  {author} {\bibfnamefont {R.~E.}\ \bibnamefont {Warburton}}, \bibinfo {author}
  {\bibfnamefont {J.}~\bibnamefont {Leach}}, \bibinfo {author} {\bibfnamefont
  {M.}~\bibnamefont {Agnew}}, \bibinfo {author} {\bibfnamefont {G.~S.}\
  \bibnamefont {Buller}}, \bibinfo {author} {\bibfnamefont {R.~W.}\
  \bibnamefont {Boyd}},\ and\ \bibinfo {author} {\bibfnamefont {M.~J.}\
  \bibnamefont {Padgett}},\ }\href {https://doi.org/10.1038/ncomms1988}
  {\bibfield  {journal} {\bibinfo  {journal} {Nature Communications}\ }\textbf
  {\bibinfo {volume} {3}},\ \bibinfo {pages} {984} (\bibinfo {year}
  {2012})}\BibitemShut {NoStop}%
\bibitem [{\citenamefont {Moreau}\ \emph {et~al.}(2014)\citenamefont {Moreau},
  \citenamefont {Devaux},\ and\ \citenamefont {Lantz}}]{Moreau2014}%
  \BibitemOpen
  \bibfield  {author} {\bibinfo {author} {\bibfnamefont {P.~A.}\ \bibnamefont
  {Moreau}}, \bibinfo {author} {\bibfnamefont {F.}~\bibnamefont {Devaux}},\
  and\ \bibinfo {author} {\bibfnamefont {E.}~\bibnamefont {Lantz}},\ }\href
  {https://doi.org/10.1103/PhysRevLett.113.160401} {\bibfield  {journal}
  {\bibinfo  {journal} {Physical Review Letters}\ }\textbf {\bibinfo {volume}
  {113}},\ \bibinfo {pages} {160401} (\bibinfo {year} {2014})}\BibitemShut
  {NoStop}%
\bibitem [{\citenamefont {Dąbrowski}\ \emph {et~al.}(2017)\citenamefont
  {Dąbrowski}, \citenamefont {Parniak},\ and\ \citenamefont
  {Wasilewski}}]{Dabrowski2017}%
  \BibitemOpen
  \bibfield  {author} {\bibinfo {author} {\bibfnamefont {M.}~\bibnamefont
  {Dąbrowski}}, \bibinfo {author} {\bibfnamefont {M.}~\bibnamefont
  {Parniak}},\ and\ \bibinfo {author} {\bibfnamefont {W.}~\bibnamefont
  {Wasilewski}},\ }\href {https://doi.org/10.1364/optica.4.000272} {\bibfield
  {journal} {\bibinfo  {journal} {Optica}\ }\textbf {\bibinfo {volume} {4}},\
  \bibinfo {pages} {272} (\bibinfo {year} {2017})}\BibitemShut {NoStop}%
\bibitem [{\citenamefont {Strekalov}\ \emph {et~al.}(1995)\citenamefont
  {Strekalov}, \citenamefont {Sergienko}, \citenamefont {Klyshko},\ and\
  \citenamefont {Shih}}]{Strekalov1995}%
  \BibitemOpen
  \bibfield  {author} {\bibinfo {author} {\bibfnamefont {D.~V.}\ \bibnamefont
  {Strekalov}}, \bibinfo {author} {\bibfnamefont {A.~V.}\ \bibnamefont
  {Sergienko}}, \bibinfo {author} {\bibfnamefont {D.~N.}\ \bibnamefont
  {Klyshko}},\ and\ \bibinfo {author} {\bibfnamefont {Y.~H.}\ \bibnamefont
  {Shih}},\ }\href {https://doi.org/10.1103/PhysRevLett.74.3600} {\bibfield
  {journal} {\bibinfo  {journal} {Physical Review Letters}\ }\textbf {\bibinfo
  {volume} {74}},\ \bibinfo {pages} {3600} (\bibinfo {year}
  {1995})}\BibitemShut {NoStop}%
\bibitem [{\citenamefont {Pittman}\ \emph {et~al.}(1995)\citenamefont
  {Pittman}, \citenamefont {Shih}, \citenamefont {Strekalov},\ and\
  \citenamefont {Sergienko}}]{Pittman1995}%
  \BibitemOpen
  \bibfield  {author} {\bibinfo {author} {\bibfnamefont {T.~B.}\ \bibnamefont
  {Pittman}}, \bibinfo {author} {\bibfnamefont {Y.~H.}\ \bibnamefont {Shih}},
  \bibinfo {author} {\bibfnamefont {D.~V.}\ \bibnamefont {Strekalov}},\ and\
  \bibinfo {author} {\bibfnamefont {A.~V.}\ \bibnamefont {Sergienko}},\ }\href
  {https://doi.org/10.1103/PhysRevA.52.R3429} {\bibfield  {journal} {\bibinfo
  {journal} {Physical Review A}\ }\textbf {\bibinfo {volume} {52}},\ \bibinfo
  {pages} {R3429} (\bibinfo {year} {1995})}\BibitemShut {NoStop}%
\bibitem [{\citenamefont {Moreau}\ \emph
  {et~al.}(2019{\natexlab{a}})\citenamefont {Moreau}, \citenamefont
  {Toninelli}, \citenamefont {Gregory},\ and\ \citenamefont
  {Padgett}}]{Moreau2019a}%
  \BibitemOpen
  \bibfield  {author} {\bibinfo {author} {\bibfnamefont {P.-A.}\ \bibnamefont
  {Moreau}}, \bibinfo {author} {\bibfnamefont {E.}~\bibnamefont {Toninelli}},
  \bibinfo {author} {\bibfnamefont {T.}~\bibnamefont {Gregory}},\ and\ \bibinfo
  {author} {\bibfnamefont {M.~J.}\ \bibnamefont {Padgett}},\ }\href
  {https://doi.org/10.1038/s42254-019-0056-0} {\bibfield  {journal} {\bibinfo
  {journal} {Nature Reviews Physics}\ }\textbf {\bibinfo {volume} {1}},\
  \bibinfo {pages} {367} (\bibinfo {year} {2019}{\natexlab{a}})}\BibitemShut
  {NoStop}%
\bibitem [{\citenamefont {Parniak}\ \emph {et~al.}(2017)\citenamefont
  {Parniak}, \citenamefont {Dąbrowski}, \citenamefont {Mazelanik},
  \citenamefont {Leszczy{\'{n}}ski}, \citenamefont {Lipka},\ and\ \citenamefont
  {Wasilewski}}]{Parniak2017}%
  \BibitemOpen
  \bibfield  {author} {\bibinfo {author} {\bibfnamefont {M.}~\bibnamefont
  {Parniak}}, \bibinfo {author} {\bibfnamefont {M.}~\bibnamefont {Dąbrowski}},
  \bibinfo {author} {\bibfnamefont {M.}~\bibnamefont {Mazelanik}}, \bibinfo
  {author} {\bibfnamefont {A.}~\bibnamefont {Leszczy{\'{n}}ski}}, \bibinfo
  {author} {\bibfnamefont {M.}~\bibnamefont {Lipka}},\ and\ \bibinfo {author}
  {\bibfnamefont {W.}~\bibnamefont {Wasilewski}},\ }\href
  {https://doi.org/10.1038/s41467-017-02366-7} {\bibfield  {journal} {\bibinfo
  {journal} {Nature Communications}\ }\textbf {\bibinfo {volume} {8}},\
  \bibinfo {pages} {2140} (\bibinfo {year} {2017})}\BibitemShut {NoStop}%
\bibitem [{\citenamefont {Dąbrowski}\ \emph {et~al.}(2018)\citenamefont
  {Dąbrowski}, \citenamefont {Mazelanik}, \citenamefont {Parniak},
  \citenamefont {Leszczy{\'{n}}ski}, \citenamefont {Lipka},\ and\ \citenamefont
  {Wasilewski}}]{Dabrowski2018}%
  \BibitemOpen
  \bibfield  {author} {\bibinfo {author} {\bibfnamefont {M.}~\bibnamefont
  {Dąbrowski}}, \bibinfo {author} {\bibfnamefont {M.}~\bibnamefont
  {Mazelanik}}, \bibinfo {author} {\bibfnamefont {M.}~\bibnamefont {Parniak}},
  \bibinfo {author} {\bibfnamefont {A.}~\bibnamefont {Leszczy{\'{n}}ski}},
  \bibinfo {author} {\bibfnamefont {M.}~\bibnamefont {Lipka}},\ and\ \bibinfo
  {author} {\bibfnamefont {W.}~\bibnamefont {Wasilewski}},\ }\href
  {https://doi.org/10.1103/PhysRevA.98.042126} {\bibfield  {journal} {\bibinfo
  {journal} {Physical Review A}\ }\textbf {\bibinfo {volume} {98}},\ \bibinfo
  {pages} {42126} (\bibinfo {year} {2018})}\BibitemShut {NoStop}%
\bibitem [{\citenamefont {Parniak}\ \emph {et~al.}(2019)\citenamefont
  {Parniak}, \citenamefont {Mazelanik}, \citenamefont {Leszczy{\'{n}}ski},
  \citenamefont {Lipka}, \citenamefont {Dąbrowski},\ and\ \citenamefont
  {Wasilewski}}]{Parniak2019}%
  \BibitemOpen
  \bibfield  {author} {\bibinfo {author} {\bibfnamefont {M.}~\bibnamefont
  {Parniak}}, \bibinfo {author} {\bibfnamefont {M.}~\bibnamefont {Mazelanik}},
  \bibinfo {author} {\bibfnamefont {A.}~\bibnamefont {Leszczy{\'{n}}ski}},
  \bibinfo {author} {\bibfnamefont {M.}~\bibnamefont {Lipka}}, \bibinfo
  {author} {\bibfnamefont {M.}~\bibnamefont {Dąbrowski}},\ and\ \bibinfo
  {author} {\bibfnamefont {W.}~\bibnamefont {Wasilewski}},\ }\href
  {https://doi.org/10.1103/PhysRevLett.122.063604} {\bibfield  {journal}
  {\bibinfo  {journal} {Physical Review Letters}\ }\textbf {\bibinfo {volume}
  {122}},\ \bibinfo {pages} {063604} (\bibinfo {year} {2019})}\BibitemShut
  {NoStop}%
\bibitem [{\citenamefont {Peyronel}\ \emph {et~al.}(2012)\citenamefont
  {Peyronel}, \citenamefont {Firstenberg}, \citenamefont {Liang}, \citenamefont
  {Hofferberth}, \citenamefont {Gorshkov}, \citenamefont {Pohl}, \citenamefont
  {Lukin},\ and\ \citenamefont {Vuleti{\'{c}}}}]{Peyronel2012}%
  \BibitemOpen
  \bibfield  {author} {\bibinfo {author} {\bibfnamefont {T.}~\bibnamefont
  {Peyronel}}, \bibinfo {author} {\bibfnamefont {O.}~\bibnamefont
  {Firstenberg}}, \bibinfo {author} {\bibfnamefont {Q.-Y.}\ \bibnamefont
  {Liang}}, \bibinfo {author} {\bibfnamefont {S.}~\bibnamefont {Hofferberth}},
  \bibinfo {author} {\bibfnamefont {A.~V.}\ \bibnamefont {Gorshkov}}, \bibinfo
  {author} {\bibfnamefont {T.}~\bibnamefont {Pohl}}, \bibinfo {author}
  {\bibfnamefont {M.~D.}\ \bibnamefont {Lukin}},\ and\ \bibinfo {author}
  {\bibfnamefont {V.}~\bibnamefont {Vuleti{\'{c}}}},\ }\href
  {https://doi.org/10.1038/nature11361} {\bibfield  {journal} {\bibinfo
  {journal} {Nature}\ }\textbf {\bibinfo {volume} {488}},\ \bibinfo {pages}
  {57} (\bibinfo {year} {2012})}\BibitemShut {NoStop}%
\bibitem [{\citenamefont {Bennink}\ \emph {et~al.}(2002)\citenamefont
  {Bennink}, \citenamefont {Bentley},\ and\ \citenamefont
  {Boyd}}]{Bennink2002}%
  \BibitemOpen
  \bibfield  {author} {\bibinfo {author} {\bibfnamefont {R.~S.}\ \bibnamefont
  {Bennink}}, \bibinfo {author} {\bibfnamefont {S.~J.}\ \bibnamefont
  {Bentley}},\ and\ \bibinfo {author} {\bibfnamefont {R.~W.}\ \bibnamefont
  {Boyd}},\ }\href {https://doi.org/10.1103/PhysRevLett.89.113601} {\bibfield
  {journal} {\bibinfo  {journal} {Physical Review Letters}\ }\textbf {\bibinfo
  {volume} {89}},\ \bibinfo {pages} {113601} (\bibinfo {year}
  {2002})}\BibitemShut {NoStop}%
\bibitem [{\citenamefont {Valencia}\ \emph {et~al.}(2005)\citenamefont
  {Valencia}, \citenamefont {Scarcelli}, \citenamefont {D'Angelo},\ and\
  \citenamefont {Shih}}]{Valencia2005}%
  \BibitemOpen
  \bibfield  {author} {\bibinfo {author} {\bibfnamefont {A.}~\bibnamefont
  {Valencia}}, \bibinfo {author} {\bibfnamefont {G.}~\bibnamefont {Scarcelli}},
  \bibinfo {author} {\bibfnamefont {M.}~\bibnamefont {D'Angelo}},\ and\
  \bibinfo {author} {\bibfnamefont {Y.}~\bibnamefont {Shih}},\ }\href
  {https://doi.org/10.1103/PhysRevLett.94.063601} {\bibfield  {journal}
  {\bibinfo  {journal} {Physical Review Letters}\ }\textbf {\bibinfo {volume}
  {94}},\ \bibinfo {pages} {063601} (\bibinfo {year} {2005})}\BibitemShut
  {NoStop}%
\bibitem [{\citenamefont {Aspden}\ \emph {et~al.}(2016)\citenamefont {Aspden},
  \citenamefont {Morris}, \citenamefont {He}, \citenamefont {Chen},\ and\
  \citenamefont {Padgett}}]{Aspden2016}%
  \BibitemOpen
  \bibfield  {author} {\bibinfo {author} {\bibfnamefont {R.~S.}\ \bibnamefont
  {Aspden}}, \bibinfo {author} {\bibfnamefont {P.~A.}\ \bibnamefont {Morris}},
  \bibinfo {author} {\bibfnamefont {R.}~\bibnamefont {He}}, \bibinfo {author}
  {\bibfnamefont {Q.}~\bibnamefont {Chen}},\ and\ \bibinfo {author}
  {\bibfnamefont {M.~J.}\ \bibnamefont {Padgett}},\ }\href
  {https://doi.org/10.1088/2040-8978/18/5/055204} {\bibfield  {journal}
  {\bibinfo  {journal} {Journal of Optics (United Kingdom)}\ }\textbf {\bibinfo
  {volume} {18}},\ \bibinfo {pages} {055204} (\bibinfo {year}
  {2016})}\BibitemShut {NoStop}%
\bibitem [{\citenamefont {Jack}\ \emph {et~al.}(2009)\citenamefont {Jack},
  \citenamefont {Leach}, \citenamefont {Romero}, \citenamefont {Franke-Arnold},
  \citenamefont {Ritsch-Marte}, \citenamefont {Barnett},\ and\ \citenamefont
  {Padgett}}]{Jack2009}%
  \BibitemOpen
  \bibfield  {author} {\bibinfo {author} {\bibfnamefont {B.}~\bibnamefont
  {Jack}}, \bibinfo {author} {\bibfnamefont {J.}~\bibnamefont {Leach}},
  \bibinfo {author} {\bibfnamefont {J.}~\bibnamefont {Romero}}, \bibinfo
  {author} {\bibfnamefont {S.}~\bibnamefont {Franke-Arnold}}, \bibinfo {author}
  {\bibfnamefont {M.}~\bibnamefont {Ritsch-Marte}}, \bibinfo {author}
  {\bibfnamefont {S.~M.}\ \bibnamefont {Barnett}},\ and\ \bibinfo {author}
  {\bibfnamefont {M.~J.}\ \bibnamefont {Padgett}},\ }\href
  {https://doi.org/10.1103/PhysRevLett.103.083602} {\bibfield  {journal}
  {\bibinfo  {journal} {Physical Review Letters}\ }\textbf {\bibinfo {volume}
  {103}},\ \bibinfo {pages} {083602} (\bibinfo {year} {2009})}\BibitemShut
  {NoStop}%
\bibitem [{\citenamefont {Dangelo}\ \emph {et~al.}(2005)\citenamefont
  {Dangelo}, \citenamefont {Valencia}, \citenamefont {Rubin},\ and\
  \citenamefont {Shih}}]{Dangelo2005}%
  \BibitemOpen
  \bibfield  {author} {\bibinfo {author} {\bibfnamefont {M.}~\bibnamefont
  {Dangelo}}, \bibinfo {author} {\bibfnamefont {A.}~\bibnamefont {Valencia}},
  \bibinfo {author} {\bibfnamefont {M.~H.}\ \bibnamefont {Rubin}},\ and\
  \bibinfo {author} {\bibfnamefont {Y.}~\bibnamefont {Shih}},\ }\href
  {https://doi.org/10.1103/PhysRevA.72.013810} {\bibfield  {journal} {\bibinfo
  {journal} {Physical Review A}\ }\textbf {\bibinfo {volume} {72}},\ \bibinfo
  {pages} {013810} (\bibinfo {year} {2005})}\BibitemShut {NoStop}%
\bibitem [{\citenamefont {Moreau}\ \emph
  {et~al.}(2019{\natexlab{b}})\citenamefont {Moreau}, \citenamefont
  {Toninelli}, \citenamefont {Gregory}, \citenamefont {Aspden}, \citenamefont
  {Morris},\ and\ \citenamefont {Padgett}}]{Moreau2019}%
  \BibitemOpen
  \bibfield  {author} {\bibinfo {author} {\bibfnamefont {P.~A.}\ \bibnamefont
  {Moreau}}, \bibinfo {author} {\bibfnamefont {E.}~\bibnamefont {Toninelli}},
  \bibinfo {author} {\bibfnamefont {T.}~\bibnamefont {Gregory}}, \bibinfo
  {author} {\bibfnamefont {R.~S.}\ \bibnamefont {Aspden}}, \bibinfo {author}
  {\bibfnamefont {P.~A.}\ \bibnamefont {Morris}},\ and\ \bibinfo {author}
  {\bibfnamefont {M.~J.}\ \bibnamefont {Padgett}},\ }\href
  {https://doi.org/10.1126/sciadv.aaw2563} {\bibfield  {journal} {\bibinfo
  {journal} {Science Advances}\ }\textbf {\bibinfo {volume} {5}},\ \bibinfo
  {pages} {eaaw2563} (\bibinfo {year} {2019}{\natexlab{b}})}\BibitemShut
  {NoStop}%
\bibitem [{\citenamefont {Hodgman}\ \emph {et~al.}(2019)\citenamefont
  {Hodgman}, \citenamefont {Bu}, \citenamefont {Mann}, \citenamefont
  {Khakimov},\ and\ \citenamefont {Truscott}}]{Hodgman2019}%
  \BibitemOpen
  \bibfield  {author} {\bibinfo {author} {\bibfnamefont {S.~S.}\ \bibnamefont
  {Hodgman}}, \bibinfo {author} {\bibfnamefont {W.}~\bibnamefont {Bu}},
  \bibinfo {author} {\bibfnamefont {S.~B.}\ \bibnamefont {Mann}}, \bibinfo
  {author} {\bibfnamefont {R.~I.}\ \bibnamefont {Khakimov}},\ and\ \bibinfo
  {author} {\bibfnamefont {A.~G.}\ \bibnamefont {Truscott}},\ }\href
  {https://doi.org/10.1103/PhysRevLett.122.233601} {\bibfield  {journal}
  {\bibinfo  {journal} {Physical Review Letters}\ }\textbf {\bibinfo {volume}
  {122}},\ \bibinfo {pages} {233601} (\bibinfo {year} {2019})}\BibitemShut
  {NoStop}%
\bibitem [{\citenamefont {Aspect}\ \emph {et~al.}(1981)\citenamefont {Aspect},
  \citenamefont {Grangier},\ and\ \citenamefont {Roger}}]{Aspect1981}%
  \BibitemOpen
  \bibfield  {author} {\bibinfo {author} {\bibfnamefont {A.}~\bibnamefont
  {Aspect}}, \bibinfo {author} {\bibfnamefont {P.}~\bibnamefont {Grangier}},\
  and\ \bibinfo {author} {\bibfnamefont {G.}~\bibnamefont {Roger}},\ }\href
  {https://doi.org/10.1103/PhysRevLett.47.460} {\bibfield  {journal} {\bibinfo
  {journal} {Physical Review Letters}\ }\textbf {\bibinfo {volume} {47}},\
  \bibinfo {pages} {460} (\bibinfo {year} {1981})}\BibitemShut {NoStop}%
\bibitem [{\citenamefont {Lipka}\ \emph {et~al.}(2018)\citenamefont {Lipka},
  \citenamefont {Parniak},\ and\ \citenamefont
  {Wasilewski}}]{doi:10.1063/1.5033559}%
  \BibitemOpen
  \bibfield  {author} {\bibinfo {author} {\bibfnamefont {M.}~\bibnamefont
  {Lipka}}, \bibinfo {author} {\bibfnamefont {M.}~\bibnamefont {Parniak}},\
  and\ \bibinfo {author} {\bibfnamefont {W.}~\bibnamefont {Wasilewski}},\
  }\href {https://doi.org/10.1063/1.5033559} {\bibfield  {journal} {\bibinfo
  {journal} {Applied Physics Letters}\ }\textbf {\bibinfo {volume} {112}},\
  \bibinfo {pages} {211105} (\bibinfo {year} {2018})}\BibitemShut {NoStop}%
\bibitem [{\citenamefont {Lipka}\ \emph {et~al.}(2019)\citenamefont {Lipka},
  \citenamefont {Leszczy{\'{n}}ski}, \citenamefont {Mazelanik}, \citenamefont
  {Parniak},\ and\ \citenamefont {Wasilewski}}]{Lipka2019}%
  \BibitemOpen
  \bibfield  {author} {\bibinfo {author} {\bibfnamefont {M.}~\bibnamefont
  {Lipka}}, \bibinfo {author} {\bibfnamefont {A.}~\bibnamefont
  {Leszczy{\'{n}}ski}}, \bibinfo {author} {\bibfnamefont {M.}~\bibnamefont
  {Mazelanik}}, \bibinfo {author} {\bibfnamefont {M.}~\bibnamefont {Parniak}},\
  and\ \bibinfo {author} {\bibfnamefont {W.}~\bibnamefont {Wasilewski}},\
  }\href {https://doi.org/10.1103/PhysRevApplied.11.034049} {\bibfield
  {journal} {\bibinfo  {journal} {Physical Review Applied}\ }\textbf {\bibinfo
  {volume} {11}},\ \bibinfo {pages} {034049} (\bibinfo {year}
  {2019})}\BibitemShut {NoStop}%
\bibitem [{\citenamefont {Clauser}\ \emph {et~al.}(1969)\citenamefont
  {Clauser}, \citenamefont {Horne}, \citenamefont {Shimony},\ and\
  \citenamefont {Holt}}]{Clauser1969}%
  \BibitemOpen
  \bibfield  {author} {\bibinfo {author} {\bibfnamefont {J.~F.}\ \bibnamefont
  {Clauser}}, \bibinfo {author} {\bibfnamefont {M.~A.}\ \bibnamefont {Horne}},
  \bibinfo {author} {\bibfnamefont {A.}~\bibnamefont {Shimony}},\ and\ \bibinfo
  {author} {\bibfnamefont {R.~A.}\ \bibnamefont {Holt}},\ }\href
  {https://doi.org/10.1103/PhysRevLett.23.880} {\bibfield  {journal} {\bibinfo
  {journal} {Physical Review Letters}\ }\textbf {\bibinfo {volume} {23}},\
  \bibinfo {pages} {880} (\bibinfo {year} {1969})}\BibitemShut {NoStop}%
\bibitem [{\citenamefont {Clauser}\ and\ \citenamefont
  {Horne}(1974)}]{Clauser1974}%
  \BibitemOpen
  \bibfield  {author} {\bibinfo {author} {\bibfnamefont {J.~F.}\ \bibnamefont
  {Clauser}}\ and\ \bibinfo {author} {\bibfnamefont {M.~A.}\ \bibnamefont
  {Horne}},\ }\href {https://doi.org/10.1103/PhysRevD.10.526} {\bibfield
  {journal} {\bibinfo  {journal} {Physical Review D}\ }\textbf {\bibinfo
  {volume} {10}},\ \bibinfo {pages} {526} (\bibinfo {year} {1974})}\BibitemShut
  {NoStop}%
\bibitem [{\citenamefont {Lipka}\ \emph {et~al.}(2021)\citenamefont {Lipka},
  \citenamefont {Mazelanik}, \citenamefont {Leszczyński}, \citenamefont
  {Wasilewski},\ and\ \citenamefont {Parniak}}]{Lipka2021}%
  \BibitemOpen
  \bibfield  {author} {\bibinfo {author} {\bibfnamefont {M.}~\bibnamefont
  {Lipka}}, \bibinfo {author} {\bibfnamefont {M.}~\bibnamefont {Mazelanik}},
  \bibinfo {author} {\bibfnamefont {A.}~\bibnamefont {Leszczyński}}, \bibinfo
  {author} {\bibfnamefont {W.}~\bibnamefont {Wasilewski}},\ and\ \bibinfo
  {author} {\bibfnamefont {M.}~\bibnamefont {Parniak}},\ }\href
  {https://doi.org/10.1038/s42005-021-00551-1} {\bibfield  {journal} {\bibinfo
  {journal} {Communications Physics}\ }\textbf {\bibinfo {volume} {4}},\
  \bibinfo {pages} {46} (\bibinfo {year} {2021})}\BibitemShut {NoStop}%
\bibitem [{\citenamefont {Hardy}\ and\ \citenamefont
  {Shapiro}(2013)}]{Hardy2013}%
  \BibitemOpen
  \bibfield  {author} {\bibinfo {author} {\bibfnamefont {N.~D.}\ \bibnamefont
  {Hardy}}\ and\ \bibinfo {author} {\bibfnamefont {J.~H.}\ \bibnamefont
  {Shapiro}},\ }\href {https://doi.org/10.1103/PhysRevA.87.023820} {\bibfield
  {journal} {\bibinfo  {journal} {Physical Review A}\ }\textbf {\bibinfo
  {volume} {87}},\ \bibinfo {pages} {023820} (\bibinfo {year}
  {2013})}\BibitemShut {NoStop}%
\bibitem [{\citenamefont {Shapiro}(2020)}]{Shapiro2020}%
  \BibitemOpen
  \bibfield  {author} {\bibinfo {author} {\bibfnamefont {J.~H.}\ \bibnamefont
  {Shapiro}},\ }\href {https://doi.org/10.1109/MAES.2019.2957870} {\bibfield
  {journal} {\bibinfo  {journal} {IEEE Aerospace and Electronic Systems
  Magazine}\ }\textbf {\bibinfo {volume} {35}},\ \bibinfo {pages} {8} (\bibinfo
  {year} {2020})}\BibitemShut {NoStop}%
\bibitem [{\citenamefont {Pirandola}\ \emph {et~al.}(2018)\citenamefont
  {Pirandola}, \citenamefont {Bardhan}, \citenamefont {Gehring}, \citenamefont
  {Weedbrook},\ and\ \citenamefont {Lloyd}}]{Pirandola2018}%
  \BibitemOpen
  \bibfield  {author} {\bibinfo {author} {\bibfnamefont {S.}~\bibnamefont
  {Pirandola}}, \bibinfo {author} {\bibfnamefont {B.~R.}\ \bibnamefont
  {Bardhan}}, \bibinfo {author} {\bibfnamefont {T.}~\bibnamefont {Gehring}},
  \bibinfo {author} {\bibfnamefont {C.}~\bibnamefont {Weedbrook}},\ and\
  \bibinfo {author} {\bibfnamefont {S.}~\bibnamefont {Lloyd}},\ }\href
  {https://doi.org/10.1038/s41566-018-0301-6} {\bibfield  {journal} {\bibinfo
  {journal} {Nature Photonics}\ }\textbf {\bibinfo {volume} {12}},\ \bibinfo
  {pages} {724} (\bibinfo {year} {2018})}\BibitemShut {NoStop}%
\bibitem [{\citenamefont {Gregory}\ \emph {et~al.}(2020)\citenamefont
  {Gregory}, \citenamefont {Moreau}, \citenamefont {Toninelli},\ and\
  \citenamefont {Padgett}}]{Gregory2020}%
  \BibitemOpen
  \bibfield  {author} {\bibinfo {author} {\bibfnamefont {T.}~\bibnamefont
  {Gregory}}, \bibinfo {author} {\bibfnamefont {P.-A.}\ \bibnamefont {Moreau}},
  \bibinfo {author} {\bibfnamefont {E.}~\bibnamefont {Toninelli}},\ and\
  \bibinfo {author} {\bibfnamefont {M.~J.}\ \bibnamefont {Padgett}},\ }\href
  {https://doi.org/10.1126/sciadv.aay2652} {\bibfield  {journal} {\bibinfo
  {journal} {Science Advances}\ }\textbf {\bibinfo {volume} {6}},\ \bibinfo
  {pages} {eaay2652} (\bibinfo {year} {2020})}\BibitemShut {NoStop}%
\bibitem [{\citenamefont {Defienne}\ \emph {et~al.}(2018)\citenamefont
  {Defienne}, \citenamefont {Reichert},\ and\ \citenamefont
  {Fleischer}}]{Defienne2018}%
  \BibitemOpen
  \bibfield  {author} {\bibinfo {author} {\bibfnamefont {H.}~\bibnamefont
  {Defienne}}, \bibinfo {author} {\bibfnamefont {M.}~\bibnamefont {Reichert}},\
  and\ \bibinfo {author} {\bibfnamefont {J.~W.}\ \bibnamefont {Fleischer}},\
  }\href {https://doi.org/10.1103/PhysRevLett.121.233601} {\bibfield  {journal}
  {\bibinfo  {journal} {Physical Review Letters}\ }\textbf {\bibinfo {volume}
  {121}},\ \bibinfo {pages} {233601} (\bibinfo {year} {2018})}\BibitemShut
  {NoStop}%
\bibitem [{\citenamefont {Mazelanik}\ \emph {et~al.}(2019)\citenamefont
  {Mazelanik}, \citenamefont {Parniak}, \citenamefont {Leszczy{\'{n}}ski},
  \citenamefont {Lipka},\ and\ \citenamefont {Wasilewski}}]{Mazelanik2019}%
  \BibitemOpen
  \bibfield  {author} {\bibinfo {author} {\bibfnamefont {M.}~\bibnamefont
  {Mazelanik}}, \bibinfo {author} {\bibfnamefont {M.}~\bibnamefont {Parniak}},
  \bibinfo {author} {\bibfnamefont {A.}~\bibnamefont {Leszczy{\'{n}}ski}},
  \bibinfo {author} {\bibfnamefont {M.}~\bibnamefont {Lipka}},\ and\ \bibinfo
  {author} {\bibfnamefont {W.}~\bibnamefont {Wasilewski}},\ }\href
  {https://doi.org/10.1038/s41534-019-0136-0} {\bibfield  {journal} {\bibinfo
  {journal} {npj Quantum Information}\ }\textbf {\bibinfo {volume} {5}},\
  \bibinfo {pages} {22} (\bibinfo {year} {2019})}\BibitemShut {NoStop}%
\bibitem [{\citenamefont {Defienne}\ \emph {et~al.}(2019)\citenamefont
  {Defienne}, \citenamefont {Reichert}, \citenamefont {Fleischer},\ and\
  \citenamefont {Faccio}}]{Defienne2019}%
  \BibitemOpen
  \bibfield  {author} {\bibinfo {author} {\bibfnamefont {H.}~\bibnamefont
  {Defienne}}, \bibinfo {author} {\bibfnamefont {M.}~\bibnamefont {Reichert}},
  \bibinfo {author} {\bibfnamefont {J.~W.}\ \bibnamefont {Fleischer}},\ and\
  \bibinfo {author} {\bibfnamefont {D.}~\bibnamefont {Faccio}},\ }\href
  {https://doi.org/10.1126/sciadv.aax0307} {\bibfield  {journal} {\bibinfo
  {journal} {Science Advances}\ }\textbf {\bibinfo {volume} {5}},\ \bibinfo
  {pages} {eaax0307} (\bibinfo {year} {2019})}\BibitemShut {NoStop}%
\bibitem [{\citenamefont {Malik}\ \emph {et~al.}(2012)\citenamefont {Malik},
  \citenamefont {Maga{\~{n}}a-Loaiza},\ and\ \citenamefont {Boyd}}]{Malik2012}%
  \BibitemOpen
  \bibfield  {author} {\bibinfo {author} {\bibfnamefont {M.}~\bibnamefont
  {Malik}}, \bibinfo {author} {\bibfnamefont {O.~S.}\ \bibnamefont
  {Maga{\~{n}}a-Loaiza}},\ and\ \bibinfo {author} {\bibfnamefont {R.~W.}\
  \bibnamefont {Boyd}},\ }\href {https://doi.org/10.1063/1.4770298} {\bibfield
  {journal} {\bibinfo  {journal} {Applied Physics Letters}\ }\textbf {\bibinfo
  {volume} {101}},\ \bibinfo {pages} {241103} (\bibinfo {year}
  {2012})}\BibitemShut {NoStop}%
\bibitem [{\citenamefont {Yao}\ \emph {et~al.}(2018)\citenamefont {Yao},
  \citenamefont {Liu}, \citenamefont {You}, \citenamefont {Wang}, \citenamefont
  {Feng}, \citenamefont {Liu}, \citenamefont {Cui}, \citenamefont {Huang},\
  and\ \citenamefont {Zhang}}]{Yao2018}%
  \BibitemOpen
  \bibfield  {author} {\bibinfo {author} {\bibfnamefont {X.}~\bibnamefont
  {Yao}}, \bibinfo {author} {\bibfnamefont {X.}~\bibnamefont {Liu}}, \bibinfo
  {author} {\bibfnamefont {L.}~\bibnamefont {You}}, \bibinfo {author}
  {\bibfnamefont {Z.}~\bibnamefont {Wang}}, \bibinfo {author} {\bibfnamefont
  {X.}~\bibnamefont {Feng}}, \bibinfo {author} {\bibfnamefont {F.}~\bibnamefont
  {Liu}}, \bibinfo {author} {\bibfnamefont {K.}~\bibnamefont {Cui}}, \bibinfo
  {author} {\bibfnamefont {Y.}~\bibnamefont {Huang}},\ and\ \bibinfo {author}
  {\bibfnamefont {W.}~\bibnamefont {Zhang}},\ }\href
  {https://doi.org/10.1103/PhysRevA.98.063816} {\bibfield  {journal} {\bibinfo
  {journal} {Physical Review A}\ }\textbf {\bibinfo {volume} {98}},\ \bibinfo
  {pages} {063816} (\bibinfo {year} {2018})}\BibitemShut {NoStop}%
\bibitem [{\citenamefont {Aspden}\ \emph {et~al.}(2013)\citenamefont {Aspden},
  \citenamefont {Tasca}, \citenamefont {Boyd},\ and\ \citenamefont
  {Padgett}}]{Aspden2013}%
  \BibitemOpen
  \bibfield  {author} {\bibinfo {author} {\bibfnamefont {R.~S.}\ \bibnamefont
  {Aspden}}, \bibinfo {author} {\bibfnamefont {D.~S.}\ \bibnamefont {Tasca}},
  \bibinfo {author} {\bibfnamefont {R.~W.}\ \bibnamefont {Boyd}},\ and\
  \bibinfo {author} {\bibfnamefont {M.~J.}\ \bibnamefont {Padgett}},\ }\href
  {https://doi.org/10.1088/1367-2630/15/7/073032} {\bibfield  {journal}
  {\bibinfo  {journal} {New Journal of Physics}\ }\textbf {\bibinfo {volume}
  {15}},\ \bibinfo {pages} {073032} (\bibinfo {year} {2013})}\BibitemShut
  {NoStop}%
\bibitem [{\citenamefont {Aspden}\ \emph {et~al.}(2015)\citenamefont {Aspden},
  \citenamefont {Gemmell}, \citenamefont {Morris}, \citenamefont {Tasca},
  \citenamefont {Mertens}, \citenamefont {Tanner}, \citenamefont {Kirkwood},
  \citenamefont {Ruggeri}, \citenamefont {Tosi}, \citenamefont {Boyd},
  \citenamefont {Buller}, \citenamefont {Hadfield},\ and\ \citenamefont
  {Padgett}}]{Aspden2015}%
  \BibitemOpen
  \bibfield  {author} {\bibinfo {author} {\bibfnamefont {R.~S.}\ \bibnamefont
  {Aspden}}, \bibinfo {author} {\bibfnamefont {N.~R.}\ \bibnamefont {Gemmell}},
  \bibinfo {author} {\bibfnamefont {P.~A.}\ \bibnamefont {Morris}}, \bibinfo
  {author} {\bibfnamefont {D.~S.}\ \bibnamefont {Tasca}}, \bibinfo {author}
  {\bibfnamefont {L.}~\bibnamefont {Mertens}}, \bibinfo {author} {\bibfnamefont
  {M.~G.}\ \bibnamefont {Tanner}}, \bibinfo {author} {\bibfnamefont {R.~A.}\
  \bibnamefont {Kirkwood}}, \bibinfo {author} {\bibfnamefont {A.}~\bibnamefont
  {Ruggeri}}, \bibinfo {author} {\bibfnamefont {A.}~\bibnamefont {Tosi}},
  \bibinfo {author} {\bibfnamefont {R.~W.}\ \bibnamefont {Boyd}}, \bibinfo
  {author} {\bibfnamefont {G.~S.}\ \bibnamefont {Buller}}, \bibinfo {author}
  {\bibfnamefont {R.~H.}\ \bibnamefont {Hadfield}},\ and\ \bibinfo {author}
  {\bibfnamefont {M.~J.}\ \bibnamefont {Padgett}},\ }\href
  {https://doi.org/10.1364/OPTICA.2.001049} {\bibfield  {journal} {\bibinfo
  {journal} {Optica}\ }\textbf {\bibinfo {volume} {2}},\ \bibinfo {pages}
  {1049} (\bibinfo {year} {2015})}\BibitemShut {NoStop}%
\end{thebibliography}%
\clearpage
\newpage

\onecolumn
\section*{\sm}
\subsection*{Quantum Memory implementation}

The quantum memory is based on a pencil shaped cold (T$\approx50\,\mu\mathrm{K}$)
cloud of $^{87}\mathrm{Rb}$ atoms prepared in a 3D magnetooptical
trap operating at around $400\,\mathrm{Hz}$ repetition rate. The
memory protocol utilizes two atomic ground states $|g\rangle=|5^{2}S_{1/2},F\text{\LyXThinSpace}=\text{\LyXThinSpace}1\LyXThinSpace\rangle$
and $|h\rangle=|5^{2}S_{1/2},F\text{\LyXThinSpace}=\text{\LyXThinSpace}2\rangle$
to store \emph{idler} photons in a form of collective atomic excitations
known as spin waves {[}see also Fig. 1(c) of the main text{]}.
The spin waves are generated together with \emph{signal} photons in
a write-out process for which a strong $150\,\mathrm{ns}$ long write
(w) laser pulse red detuned by $30\,\mathrm{MHz}$ from $|g\rangle\to|e_{w}\rangle=|5^{2}P_{3/2},F\text{\LyXThinSpace}=\text{\LyXThinSpace}2\rangle$
transition is employed. To read-out the \emph{idler} photon we use
a strong read (r) laser pulse ($150\,\mathrm{ns}$) resonant with
$|h\rangle\to|e_{r}\rangle=\text{\LyXThinSpace}|5^{2}P_{1/2},F\text{\LyXThinSpace}=\text{\LyXThinSpace}2\rangle$
transition. The double $\Lambda$ scheme employed in the memory protocol
allow us to efficiently filter out any stray laser light present in
either \emph{signal} or \emph{idler} beam (see \citep{Parniak2017}
for details of the filtering system). Due to the phase matching involved
in the readout process the resulting photon pairs are anti-correlated
in momenta. As we choose the write and read beams to counter propagate
through the ensemble the anti-correlation reads: $\frac{\mathbf{k}_{i}}{|\mathbf{k}_{i}|}=-\frac{\mathbf{k}_{s}}{|\mathbf{k}_{s}|}$,
where $\mathbf{k}_{s(i)}$, correspond to \emph{signal} and \emph{idler}
photon wavevector respectively. Moreover, as demonstrated previously
\citep{Dabrowski2018} the resulting photons are additionally correlated
in positions, which is a feature of the EPR state.

\subsection*{CHSH inequality in a hybrid scenario}

To perform the Bell test in our hybrid scenario, we use a variation
of the CHSH inequality adjusted for our correlation function defined
(neglecting the wavevector dependence) as follows:

\begin{equation}
\mathcal{C}^{\theta_{s},\theta_{i}}=\frac{n_{+}^{\theta_{s},\theta_{i}}-n_{-}^{\theta_{s},\theta_{i}}}{n_{+}^{\theta_{s},\theta_{i}}+n_{-}^{\theta_{s},\theta_{i}}},\label{eq:corrf-sm}
\end{equation}

where $n_{\pm}^{\theta_{s},\theta_{i}}$ denotes the number of registered
\emph{signal}-\emph{idler} coincidences at each port ($+$ or $-$)
of the ``dual channel'' polarizer. To derive the $S$ parameter
for such correlation function we follow the derivation of the single
channel inequality \citep{Clauser1974} with assumptions adapted to
our hybrid scenario. For the convenience let us denote the \emph{signal}
and \emph{idler} polarizer orientations as $a\,(a')$ and $b\,(b')$.
Then using the non enhancement assumption \citep{Clauser1974} we
can write the four inequalities for \emph{signal} and \emph{idler}
photon detection: 
\begin{gather}
0\leq p_{s}^{a}(\lambda)\leq p_{s}^{\infty}(\lambda),\label{eq:ps_assum}\\
0\leq p_{s}^{a'}(\lambda)\leq p_{s}^{\infty}(\lambda),\label{eq:psp_assum}\\
0\leq p_{i,\pm}^{b}(\lambda)\leq p_{i,+}^{b}(\lambda)+p_{i,-}^{b}(\lambda),\label{eq:pi_assum}\\
0\leq p_{i,\pm}^{b'}(\lambda)\leq p_{i,+}^{b'}(\lambda)+p_{i,-}^{b'}(\lambda),\label{eq:pip_assum}
\end{gather}
where $\lambda$ is the hidden variable defining the quantum state
of the source and has associated probability density $\rho(\lambda)$.
Additionally, without loss of generality we can assume $p_{i,+}^{b'}(\lambda)+p_{i,-}^{b'}(\lambda)\leq p_{i,+}^{b}(\lambda)+p_{i,-}^{b}(\lambda)$
(which in the most sensible cases is just equality) to arrive with
the following expressions:
\begin{equation}
-p_{s}^{\infty}(\lambda)p_{i,+}^{b}(\lambda)\leq p_{s}^{a}(\lambda)p_{i,+}^{b}(\lambda)-p_{s}^{a}(\lambda)p_{i,+}^{b'}(\lambda)+\label{eq:ineq_p}
p_{s}^{a'}(\lambda)p_{i,+}^{b}(\lambda)+p_{s}^{a'}(\lambda)p_{i,+}^{b'}(\lambda)\leq p_{s}^{\infty}(\lambda)p_{i,-}^{b}(\lambda),
\end{equation}
\begin{equation}
-p_{s}^{\infty}(\lambda)p_{i,-}^{b}(\lambda)\leq p_{s}^{a}(\lambda)p_{i,-}^{b}(\lambda)-p_{s}^{a}(\lambda)p_{i,-}^{b'}(\lambda)+
p_{s}^{a'}(\lambda)p_{i,-}^{b}(\lambda)+p_{s}^{a'}(\lambda)p_{i,-}^{b'}(\lambda)\leq p_{s}^{\infty}(\lambda)p_{i,+}^{b}(\lambda).\label{eq:ineq_m}
\end{equation}

Then, using the standard expression for the LHVT coincidence probability
$p_{\pm}^{a,b}(\lambda)=-p_{s}^{a}(\lambda)p_{i,\pm}^{b}(\lambda)$
and integrating inequalities \eqref{eq:ineq_p} and \eqref{eq:ineq_m}
with the hidden parameter distribution $\rho(\lambda)$ and finally
adding them with opposite signs we obtain:
\begin{equation}
-p_{+}^{\infty,b}-p_{-}^{\infty,b}\leq\mathcal{P}^{a,b}-\mathcal{P}^{a,b'}+\mathcal{P}^{a',b}+
\mathcal{P}^{a',b'}-\mathcal{P}^{\infty,b}\leq p_{+}^{\infty,b}+p_{-}^{\infty,b},\label{eq:CHSH_prob}
\end{equation}
where $\mathcal{P}^{a,b}=p_{+}^{a,b}-p_{-}^{a,b}$ is the probability
correlation function. As all of the probabilities in \eqref{eq:CHSH_prob}
are associated with \emph{signal}-\emph{idler} coincidences the inequality
can be tested without the total number of emissions being known. However,
to trustfully perform measurements needed to test this inequality
the source has to be very stable in time, especially when it comes
to the polarizer removal ($\infty$ setting). Therefore, to perform
the Bell test in our case we use the standard fair sampling assumption
and take the coincidences correlation function \eqref{eq:corrf-sm}
as a fair estimate of the probability correlation function $\mathcal{P}^{\theta_{s},\theta_{i}}=\mathcal{C}^{\theta_{s},\theta_{i}}/2$
for any pair of measurements settings $(\theta_{s},\theta_{i})$ excluding
the special case with the \emph{signal} polarizer removed, for which
$\mathcal{P}^{\infty,\theta_{i}}=\mathcal{C}^{\infty,\theta_{i}}$.
With these assumptions $p_{+}^{\infty,b}+p_{-}^{\infty,b}=1$ and
the inequality \eqref{eq:CHSH_prob} finally becomes: 
\begin{equation}
-2\leq\mathcal{C}^{\theta_{s},\theta_{i}}-\mathcal{C}^{\theta_{s},\theta_{i}^{\prime}}+\mathcal{C}^{\theta_{s}^{\prime},\theta_{i}}+\mathcal{C}^{\theta_{s}^{\prime},\theta_{i}^{\prime}}-2\mathcal{C}^{\infty,\theta_{i}}\leq2.\label{eq:CHSH-sm}
\end{equation}

Further, when $\mathcal{C}^{\theta_{s},\theta_{i}}=\mathcal{C}(\phi=|\theta_{s}-\theta_{i}|)$
and we choose $\theta_{s}$, $\theta_{s}^{\prime}$, $\theta_{i}$,
$\theta_{i}^{\prime}$ to satisfy $|\theta_{s}-\theta_{i}|=|\theta_{s}^{\prime}-\theta_{i}|=|\theta_{s}^{\prime}-\theta_{i}^{\prime}|=\frac{1}{3}|\theta_{s}-\theta_{i}^{\prime}|=\phi$
the inequality \eqref{eq:CHSH-sm} becomes: 
\begin{equation}
-2\leq3\mathcal{C}(\phi)-\mathcal{C}(3\phi)-2\mathcal{C}^{\infty}(\phi)\leq2,\label{eq:CHSH_phi_sm}
\end{equation}
which is the correlation function equivalent of the inequality derived
by Freedman for probabilities \citep{Freedman1972}.

\subsection*{Visibility limitations}

Here we summarize all factors contributing to reduced visibility in correlation function fringes.

First of all, the ultimate visibility
(assuming perfect optical components) is limited by the purity of
the generated photons pairs. Any stray noise present in the \emph{idler}
and indirectly \emph{signal} arm\emph{ }will appear as a nonzero floor
on the acquired ghost images. As the noise comes from uncorrelated
photon pairs it will appear also outside the region determined by
the \emph{signal} observation region (limited by the bucket detector
sensitivity area). Therefore, by looking outside the correlation region
we can get the average amount of the noise present $\bar{n}_{\mathrm{bckg}}$
and then estimate the ultimate visibility as: $\mathcal{V}_{\mathrm{ult}}=(\bar{n}_{\mathrm{sig}}-\bar{n}_{\mathrm{bckg}})/(\bar{n}_{\mathrm{sig}}+\bar{n}_{\mathrm{bckg}})$,
where $\bar{n}_{\mathrm{sig}}=\nicefrac{1}{\mathcal{A}_{\circ}}\int(n_{+}^{\infty,0}(\bk_{i})+n_{+}^{\infty,0}(\bk_{i}))\mathrm{d}\bk_{i}$
is average number of registered coincidences in the circular area
$\mathcal{A}_{\circ}$. In our case the ultimate visibility amounts
about $89\%$. It is also noteworthy that this formula can be rewritten
using Glauber second order correlation function of \emph{signal} and
\emph{idler} pairs, as $g_{si}^{(2)}\simeq\bar{n}_{\mathrm{tot}}/\bar{n}_{\mathrm{bckg}}\approx16.5$.
The imperfections of the optical components, including mirrors, polarizers
and waveplates lower the achievable visibility by a factor of $0.97^{2}$.
Additionally, finite correlation strength characterized by $\kappa$
limits the visibility for fast varying phase profiles in the \emph{signal}
arm, however as this phase slope in our case is low this effect has
minimal contribution (less than $1\%$) to the total visibility. Finally,
the remaining drop (by a factor equivalent to about $97\%$) is most
probably caused by the residual mismatch of the wavevector shift $\delta\bk$
between the two MZIs. 

The next sections of this \sm\ treat the details of various effects impacting the visbility.
\subsection*{Noise and photon pairs purity}

The ultimate visibility that could be achieved in our setup when ideal
optical components are used is limited by purity of the generated
photon pair state. As the \emph{signal}-\emph{idler} pair generation
process is purely probabilistic there always is a chance of multiple-pair
generation in one shot. The higher order pairs similarly as spurious
noise photons coming from the read laser leakage and I-CMOS sensor
dark counts will appear at the acquired images as a constant noise
floor, present even outside the correlation region determined by the
bucked detector area. Therefore, by looking outside the correlation
circle we can estimate the total amount of the noise present and hence
calculate the ultimate visibility by comparing the noise level with
the correlated coincidences. 

\begin{figure*}[t]
	\includegraphics[width=1\textwidth]{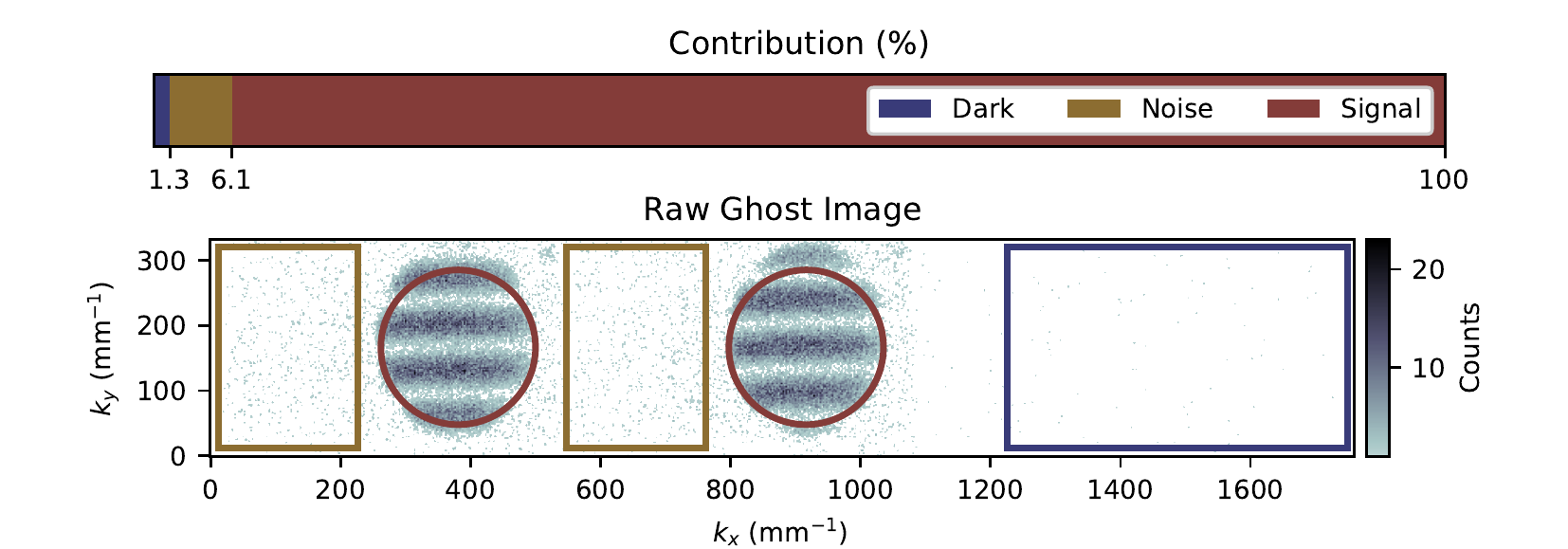}
	
	\caption{The percentage contributions of the noise to the observed signal.
		The bar represents the average value taken from the appropriate region
		in the ghost image. The dark counts are taken from the region of the
		I-CMOS camera frame that is not illuminated by any optical field (purple
		frame). The noise counts are taken from regions (golden frame) outside
		the ghost-imaged circular bucked detector area, these includes also
		the dark counts as they are present uniformly over whole frame. The
		signal counts, that include also the noise counts are taken from the
		circular regions inside the virtually-visible bucket detector area
		(red circles). The wavevector variables are only for scale purposes
		and they (in absolute sense) do not refer to the true wavevector values.\label{fig:Noise}}
\end{figure*}

In Fig. \eqref{fig:Noise} we present the percentage noise and signal
values taken from measurement with $(0,0)$ analyzer settings. From
these we estimate the ultimate visibility to be $\mathcal{V}_{\mathrm{ult}}=(\bar{n}_{\mathrm{sig}}-\bar{n}_{\mathrm{bckg}})/(\bar{n}_{\mathrm{sig}}+\bar{n}_{\mathrm{bckg}})\approx88.6\%$.

The quality of the \emph{signal}-\emph{idler} pairs can be characterized
by Glauber second order correlation function defined as 
\begin{equation}
g_{si}^{2}=\frac{\langle n_{s}n_{i}\rangle}{\bar{n}_{s}\bar{n}_{i}},\label{eq:g2_def}
\end{equation}
where $\langle n_{s}n_{i}\rangle$ is the average number of coincidences
and $\bar{n}_{s(i)}$ represents average number of \emph{s}(\emph{i})-\emph{signal}(\emph{idler})
photons. As we are working with low pair generation probability $p$
we can approximate the above averages with proper probabilities. Let
us we denote the total transition-and-detection efficiencies as $\chi_{s}$,
$\chi_{i}$ and introduce the two spurious noise probabilities $\zeta_{s}$
,$\zeta_{s}$ for \emph{s} and \emph{i }arm respectively. Then, we
can write the coincidence probability by analyzing the possible events
up to the $\mathcal{O}(p^{2})$ order: 
\begin{equation}
p_{si}=p\chi_{s}\chi_{i}+p\chi_{s}\zeta_{i}+p\chi_{i}\zeta_{s}+\zeta_{s}\zeta_{i}+p^{2}\chi_{s}\chi_{i}.
\end{equation}
The subsequent terms have the following interpretation: $p\chi_{s}\chi_{i}$
represent the probability of generating and detecting a genuine \emph{s}-\emph{i}
pair; $p\chi_{s}\zeta_{i}$, $p\chi_{i}\zeta_{s}$ are probabilities
of coincidences between the \emph{signal} (or \emph{idler}) photon
an the noise photon in the second arm; $\zeta_{s}\zeta_{i}$ is the
probability of noise-noise coincidence, and $p^{2}\chi_{s}\chi_{i}$
is the \emph{s}-\emph{i} from two generated (and uncorrelated) pairs.
The single-photon detection probabilities read: $p_{s}=p\chi_{s}+\zeta_{s}$
and $p_{i}=p\chi_{i}+\zeta_{i}$, which, after dropping the $\zeta_{s}\zeta_{i}\ll p\chi_{s(i)}\zeta_{s(i)}$
term finally give: 
\begin{equation}
g_{si}^{2}\simeq1+\frac{1}{p+\zeta_{s}/\chi_{s}+\zeta_{i}/\chi_{i}},
\end{equation}

which for $\bar{n}_{s}\simeq p_{s}$ transform into:
\begin{equation}
g_{si}^{2}\simeq1+\frac{1}{\bar{n}_{s}/\chi_{w}+\zeta_{i}/\chi_{i}}.\label{eq:g2_mod}
\end{equation}

Additionally, as mentioned earlier the genuine coincidences appear
on the camera frame only in region limited by the bucket detector
area, hence, by looking outside this region we can estimate the denominator
of \eqref{eq:g2_def} as $\bar{n}_{s}\bar{n}_{i}\simeq p_{s}p_{i}=p_{si}-p\chi_{s}\chi_{i}$
and arrive at $g_{si}^{(2)}\simeq\bar{n}_{\mathrm{tot}}/\bar{n}_{\mathrm{bckg}}$,
which in our case (Fig. \ref{fig:Noise}) amounts to $g_{si}^{(2)}\approx16.5$.

\begin{figure}[t]
	\centering
	\includegraphics[width=0.56\textwidth]{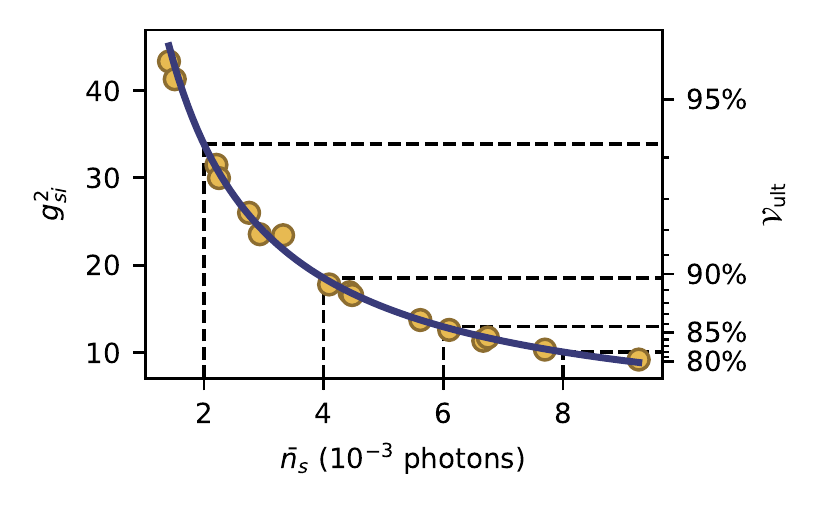}
	
	\caption{Measured second order correlation function $g_{si}^{2}$ between \emph{signal}
		and \emph{idler} photons as a function of mean \emph{signal} photon
		number $\bar{n}_{s}$. The solid curve represents a simple model fitted
		to the data. The ultimate visibility $\mathcal{V}_{\mathrm{ult}}$
		corresponding to the measured \emph{$g_{si}^{(2)}$} function values
		is present on the second vertical axis.\label{fig:g2_SM}}
\end{figure}

In Fig. \ref{fig:g2_SM} we present additional measurement of the
$g_{si}^{2}$ function for different \emph{signal }photon mean numbers.
The solid curve is fit of \eqref{eq:g2_mod} to the data with $\chi_{s}=(7.5\pm0.1)\%$
and $\zeta_{i}/\chi_{i}=(4\pm2)\permil$. The measurement for the
GI experiment were performed around $\bar{n}_{s}\approx4\times10^{-3}$,
corresponding to $p\approx0.05$.

\subsection*{Correlation strength $\kappa$ and its impact on visibility}

\begin{figure*}[t]
	\includegraphics[width=1\textwidth]{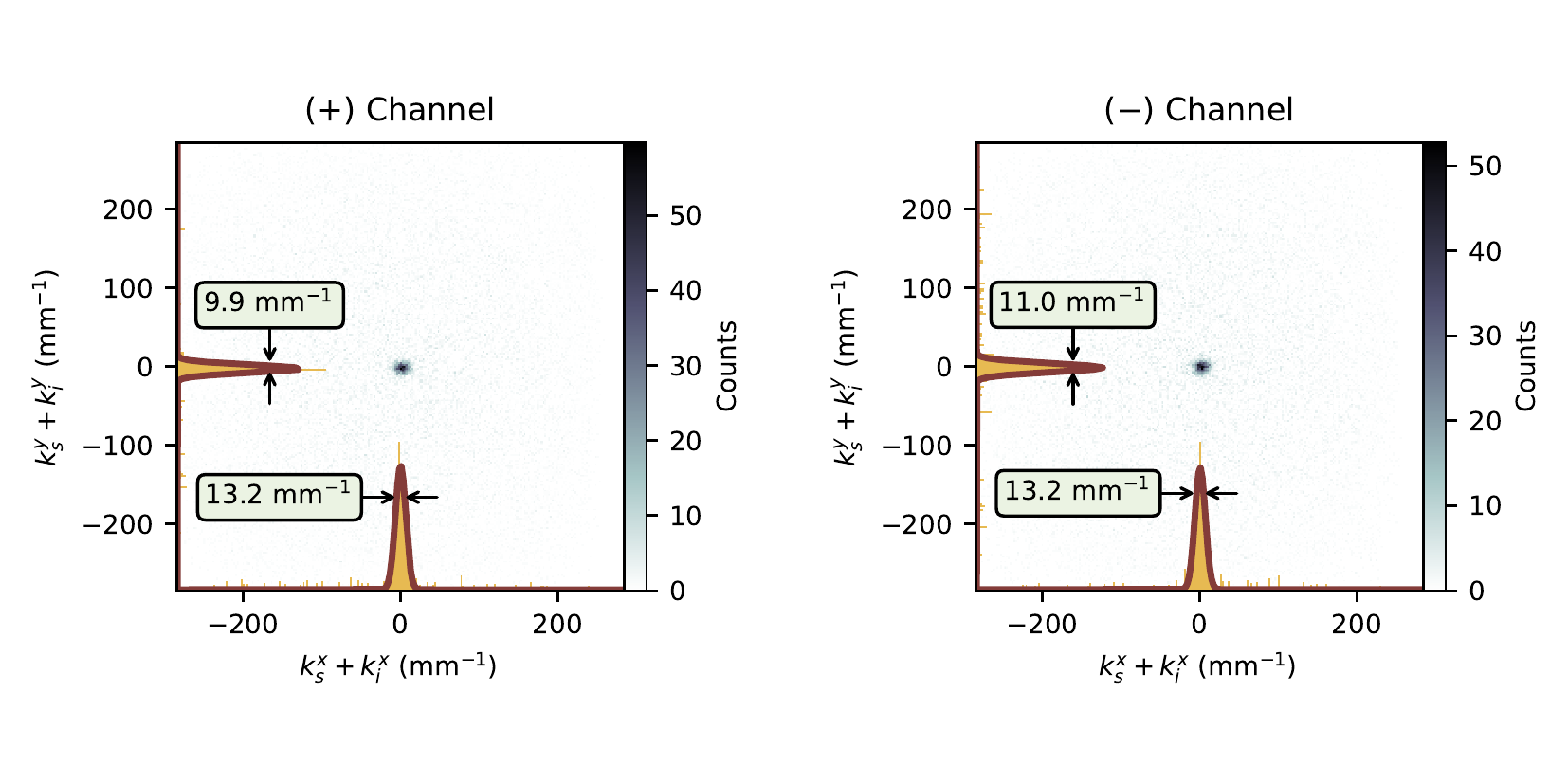}\caption{Noise-subtracted\emph{ signal}-\emph{idler} coincidences in the center
		of mass coordinates measured with camera. The solid lines represent
		the cross sections through the maximum of the fitted theoretical probability
		distribution $|\psi_{\protect\epr}(\protect\bk_{s},\protect\bk_{i})|^{2}$.
		The labels mark the $2\kappa$ widths of the distributions. \label{fig:kappa_corr}}
\end{figure*}

The generated biphoton wavefunction $\psi_{\epr}(\mathbf{k}_{s},\mathbf{k}_{i})$
is characterized by near-field ($\sigma$) and far-field ($\kappa$)
correlations strengths. In our case, since we are working with wavevector
correlations, the $\kappa$ parameter is of the most importance. It
limits, the achievable resolution and in the case of the Bell-correlations
demonstration it leads to the drop of the visibility of observed quantum
fringes. The visibility reduction due to finite $\kappa$ in this
case can be estimated by evaluating the integral (see Eq. (5)
of the main article) $p_{\pm}(\bk_{i})=\int|\tilde{\psi}_{\epr}(\bk_{s}-\delta\mathbf{k},\bk_{i})|^{2}\cos^{2}\left(\frac{\varphi(\bk_{s})-\varphi(\bk_{i})}{2}\right)\mathrm{d}\bk_{s}$
for a linearly varying $\varphi(\bk_{s})$, which without loss of
generality can by assumed to vary only in the $\hat{x}$ direction,
$\varphi(\bk_{s})=\alpha k_{x}$. Additionally, as mentioned in the
main article the wavevector-space field of view is limited by technical
constraints rather than the spread $\sigma$ of the generated state
and thus we take $\sigma\to0.$ In this case, the integral gives $(1+e^{-\alpha^{2}\kappa^{2}/2}\cos(\varphi_{s}(-\bk_{i})-\varphi(\bk_{i}))/2$,
where we can identify the visibility to be $\mathcal{V}_{\kappa}=e^{-\alpha^{2}\kappa^{2}/2}$.
To estimate the correlation strength $\kappa$ and thus the reduction
of visibility we performed additional measurement with the bucked
detector replaced by camera. The results in form of correlation maps
in center of mass coordinates $(\bk_{s}+\bk_{i})$ integrated over
the second direction $(\bk_{s}-\bk_{i})$ are present in Fig. \eqref{fig:kappa_corr}.
The maps are drawn for two channels ($+$ and $-$) of the beam displacer
(BD) with the noise background $\bar{n}_{s}\bar{n}_{i}$ subtracted.
The solid lines represents the fitted Gaussian curves which give estimates
of $\kappa$ in each direction. Since, the estimates of $\kappa$
have similar values as the final estimate we take the average over
the four values and obtain $\kappa=(5.9\pm0.7)\,\mathrm{mm^{-1}}$.
This, together with our \emph{signal} phase slope $\alpha\approx12.4\,\mathrm{mrad\times mm}$
gives $\mathcal{V}_{\kappa}\approx99.7\%$ which is negligible when
compared to other imperfections.

\subsection*{Visibility drop due to the MZIs $\delta\protect\bk$ mismatch}

To derive the formula for the desired Bell-like sate $|\psi_{\bel}\rangle$
we assumed that the two MZIs have the same (matched) wavevector shift
$\delta\bk$. In the experiential realization, however, these two
shifts will be always matched with finite precision $\xi\bk$. If
this is the case, the generated state can be no longed described by
the compact form, where the superposition is only at the polarization
level as in Eq. (3) of the main text. To quantify the effect of the
residual mismatch we can to use the intermediate form and add the
mismatch to the one of the parts: 
\begin{equation}
|\psi_{\bel}\rangle=\mathcal{N}\iint_{[0,\delta\bk]^{2}}(e^{i\varphi_{s}(\mathbf{k}_{s})}\psi_{\epr}(\mathbf{k}_{s},\mathbf{k}_{i}-\delta\mathbf{k})|\mathbf{k}_{s},H\rangle|\mathbf{k}_{i},V\rangle+e^{i\varphi_{i}(\mathbf{k}_{i})}\psi_{\epr}(\mathbf{k}_{s}-\delta\mathbf{k-\xi\mathbf{k}},\mathbf{k}_{i})|\mathbf{k}_{s},V\rangle|\mathbf{k}_{i},H\rangle)\mathrm{d}\mathbf{k}_{s}\mathrm{d}\mathbf{k}_{i}.\label{eq:psi_bell_mismatch_SM}
\end{equation}

Then, after plugging \eqref{eq:psi_bell_mismatch_SM} into Eq. (5) of the main text
and evaluating the integral in the $\delta\bk\gg\kappa$ limit we
obtain: 
\[
p\propto(1+e^{-\xi\bk^{2}/(8\kappa^{2})}\cos(\varphi_{s}(-\bk_{i})-\varphi_{i}(\bk_{i}))/2,
\]
where we identify the visibility as $\mathcal{V}_{\xi\bk}=e^{-\xi\bk^{2}/(8\kappa^{2})}$
which we estimate to be $97\%$ that correspond to $\xi\bk\approx0.5\kappa$.

\subsection*{Classical interference measurements}
To measure the phase profiles we replaced the \emph{bucket} detector with the camera and beam displacer as in the \emph{idler} path and performed interference measurements using classical light. The classical \emph{idler} light was generated in the memory by seeding the write-out process using additional laser beam that was injected to the memory simultaneously with the write lase pulse. The seed beam was phase-coherent with the write laser and focused in the center of the atomic cloud. To maintain the phase coherence the seed beam is derived from the write laser by frequency shifting small fraction of the laser output by $6.834\,\mathrm{GHz}$ using EOM and filtering cavity (see \cite{Parniak2019} for details of the setup). This beam was used both to measure the phase profile in the \emph{signal} arm and to generate strong atomic coherence in the memory. In the read-out process this coherence was retrieved to light that has been used to measure phase profile in the \emph{idler} arm. Each phase profile measurement consisted of running the experiment with continuously changing the measurement settings $\{\theta_{s},\theta_{i}\}\in[0,2\pi]\times[0,2\pi]$ and registering the intensity fringes $I_{+\, (-)}$ in the $+$ and $-$ measurement channels corresponding to two separate regions on the camera. The phase profiles were then retrieved using Fourier-transform based procedure \cite{Lipka2019} evaluated on the correlation function calculated as $\mathcal{C}=\frac{I_{+}-I_{-}}{I_{+}+I_{-}}$. In this procedure we first take a two-dimensional Fourier transform of the real-valued correlation frame, then we select spatial-frequency region around one of two main (identical) frequency components that are located symmetrically to the zero-frequency point. We filter out that region by setting remaining (outside this region) values to zero. Finally inverse two-dimensional Fourier transform gives us a complex signal with phase equal to the interferometer phase profile shifted by $\theta_{s\, (i)}$. We repeat the procedure for each correlation frame for both \emph{signal} and \emph{idler} arm. For each calculated complex frame we set the global phase $\theta_{s\, (i)}$ to 0 and finally we average the results. The final phase profiles are retrieved from the complex average by simply evaluating the $\arg(\cdot)$ function.
\end{document}